\documentclass[fleqn,usenatbib]{mnras}

\usepackage{newtxtext,newtxmath}

\usepackage[T1]{fontenc}

\DeclareRobustCommand{\VAN}[3]{#2}
\let\VANthebibliography\thebibliography
\def\thebibliography{\DeclareRobustCommand{\VAN}[3]{##3}\VANthebibliography}

\usepackage{graphicx}	
\usepackage{amsmath}	
\usepackage{soul}

\newcommand{\solmass}[1]{#1~M$_\odot$}
\newcommand{\sollum}[1]{#1~L$_\odot$}
\newcommand{\edit}[0]{}

\title[GI discs with improved cooling]{Revisiting gravitational instability in protostellar discs with improved radiative cooling models}

\author[A. K. Young et al.]{
Alison K. Young,$^{1,2,3,4,5}$\thanks{E-mail: a.k.young@leeds.ac.uk}
Ken Rice,$^{2,3}$
Richard Booth,$^{1}$
and Farzana Meru $^{4,5}$
\\
$^{1}$ School of Physics and Astronomy, University of Leeds, Sir William Henry Bragg Building, Woodhouse Ln, Leeds LS2 9JT, UK \\
$^{2}$SUPA, Institute for Astronomy, University of Edinburgh, The Royal Observatory, Blackford Hill, Edinburgh, EH9 3HJ, UK \\
$^{3}$Centre for Exoplanet Science, University of Edinburgh, Edinburgh, EH9 3HJ, UK \\
$^{4}$ Centre for Exoplanets and Habitability, University of Warwick, Coventry CV4 7AL, UK \\
$^{5}$ Department of Physics, University of Warwick, Coventry CV4 7AL, UK 
}

\date{Accepted XXX. Received YYY; in original form ZZZ}

\pubyear{2026}

\begin{document}
\label{firstpage}
\pagerange{\pageref{firstpage}--\pageref{lastpage}}
\maketitle

\begin{abstract}
Young discs are expected to be significantly more massive than those observed at $>1$~Myr and it is at this earliest stage that planet formation likely begins. Such massive discs may be susceptible to the gravitational instability (GI), therefore we need to determine the disc and stellar properties for which the GI is active to understand its role in early disc evolution and planet formation. Prior work has been limited by model assumptions and inaccuracies due to the complex nature of the thermodynamics of protostellar discs so we now revisit this question using an improved method to approximate radiative cooling within hydrodynamics simulations. We have explored a wide parameter space, representative of young protostellar discs of 0.1 to \solmass{1} and include irradiation from the host star. The parameters for which discs form spirals and fragment were found to differ to those obtained from earlier simulations. The outer regions of discs with radii of 50~au may be susceptible to fragmentation, meaning that GI-driven planet formation is not restricted to only the most extended discs. The additional thermal support due to stellar irradiation increases the disc mass that remains stable against GI: discs may reach up to $\gtrsim 0.4$~M$_*$ without fragmenting, providing a considerable quantity of material for building planets. 
Large scale spiral arms only developed for $M_*\lesssim$\solmass{0.3}, except in the most compact discs. Furthermore, the long-lived spiral structures that form tend to be flocculent and compact, indicating that large-scale spiral arms should not be considered a typical outcome of GI.
\end{abstract}

\begin{keywords}
hydrodynamics -- radiative transfer -- protoplanetary discs -- stars:formation
\end{keywords}



\section{Introduction}

Protostellar discs form with relatively high masses that can be a significant fraction of the mass of the host protostar, while the protostar itself has yet to reach its final mass
 \citep{jorgensen2009prosac,williams2011araa}. With such a high mass fraction (\(0.1 \lesssim m_{\rm disc}/m_* \lesssim 0.3\)), young
protostellar discs may be susceptible to the growth of the gravitational instability (GI) \citep{laughlin1994,rice2016,maureira2025}, which could accelerate planet formation via the direct collapse of fragments and/or through assisting the growth of dust grains \citep{boss1998,mayer2004,rice2006,longarini2023,rowther2024,rice2025}. 
It is at this early stage during the Class 0/I embedded phase ($< 1$~Myr) that planet formation is expected to begin \citep{nixon2018,segura-cox2020}. Therefore, we need to understand the evolution of discs subject to the GI to make progress in explaining how planet formation begins.

The stability parameter known as "Toomre's $Q$" indicates whether a disc's self-gravity could lead to the development of the gravitational instability and is given by
\begin{equation}
    Q = \frac{c_s \kappa}{\pi G \Sigma},
    \label{eq:toomreQ}
\end{equation}
\citep{toomre1964}, where $c_{\rm s}$ is the sound speed, $\kappa$ is the epicyclic frequency, and $\Sigma$ is the disc surface density. In the razor-thin disc limit the GI becomes active when $Q=1$ \citep{safronov1960,toomre1964} but for realistic discs observable effects of GI may manifest when $Q\lesssim 1.5$ \citep{goldreich1965,vandervoort1970,bertin1998}.
A disc that is susceptible to the growth of GI will tend to a quasi-steady state in which shock heating due to the GI quenches the instability \citep{paczynski1978}. This self-regulation mechanism acts as a thermostat that maintains $Q\sim1$ in the disc and allows spiral density waves to emerge that can act to transport angular momentum \citep[e.g.][]{lodatorice2004}. 
If, however, cooling is sufficiently rapid, or mass accretion sufficiently fast, $Q$ may decrease well below 1 and the disc may undergo fragmentation.
The cooling rate is typically evaluated by comparing the cooling time to the Keplerian frequency $\Omega_{\rm K}$ at a given radius in the disc \citep{gammie2001}:
\begin{equation}
    \label{eq:betacool}
    \beta_{\rm cool} = \Omega_{\rm K}  t_{\rm cool}.
\end{equation}
Sufficiently rapid cooling drives fragmentation of the disc since GI cannot - in a quasi-steady state - provide enough heating to stabilize the disc. A cooling parameter $\beta_{\rm cool} \lesssim 8$ will typically result in fragmentation \citep[e.g.][]{gammie2001,rice2011,leedham2025}.
A longer cooling timescale of $\beta_{\rm cool} \gtrsim$~10--20 maintains the quasi-steady state \citep{kratter2016}.

Of course, discs are irradiated by their host star(s), which significantly affects the thermal balance and raises the mid-plane temperature. The degree to which stellar irradiation dominates the thermal balance has been studied with various analytical models and numerical simulations. With stellar irradiation, discs can reach higher masses before becoming unstable \citep{matzner2005,cai2008,meru2010,forgan2013,cadman2020a,haworth2020}. However, once the GI emerges, the maximum value of gravitational stress ($\alpha_{\rm grav}$) that can be sustained is lower for an irradiated disc and the critical value of $\beta_{\rm cool}$ necessary for fragmentation is reduced by around a factor of two with irradiation \citep{rice2011}. The result is that a long-lived self-regulated state is unlikely to be reached in irradiated discs since GI shock heating and viscous heating comprise a smaller fraction of the thermal balance. Hence, when the system is susceptible to the GI, they are probably insufficient to prevent rapid growth of the instability \citep{kratter2011,rowther2024b,leedham2025}.

The disc temperature and the cooling rate are functions of radius, with lower temperatures and lower values of $\beta_{\rm cool}$ reached further from the host star. Consequently, it was expected that the warm inner disc is stable and that fragmentation is only possible in the outer regions beyond $\sim 70$~au \citep{rafikov2005,clarke2009,rafikov2009,linkratter2016}. Under these conditions, disc fragmentation due to GI is unlikely to form many planetary-mass objects \citep{kratter2010,rice2015,forgan2018,schib2025}. However, once magnetic fields are included, the formation of sub-Jovian mass planets may be possible \citep{deng2021}
There are indications, too, that fragmentation may be possible a little closer to the host star than 70~au \citep{mercer2020,leedham2025}, and as close as $\sim 30$~au when dust grain growth is considered \citep{lee2025}. Moreover, \citet{meru2015} demonstrated that the gas motions in the disc triggered by the local collapse of a fragment may in turn cause fragmentation within 50~au.

The evolution of gravitationally unstable discs sensitively depends on the balance of heating and cooling \citep{gammie2001}. In most of the early work, the cooling rate was imposed on the model disc by means of a fixed value of $\beta_{\rm cool}$, the ratio of cooling time to orbital period (see eq.~\ref{eq:betacool}). {\edit A constant $\beta_{\rm cool}$ is a poor approximation to the cooling in protostellar discs \citep{mercer2018} and so some simulations have employed a radial-dependent $\beta_{\rm cool}=\beta_{\rm cool}(R)$ \citep{rowther2020} or temperature-dependent $\beta_{\rm cool}=\beta_{\rm cool}(T)$ \citep{cossins2010opacity}. These approaches capture better the variation in cooling times throughout the disc and tend to produce discs that are more susceptible to fragmentation.}

The ``$\beta$-cooling'' method has proved a useful approximation for studying the evolution of massive discs but further progress with $\beta$-cooling simulations is limited because the cooling rate cannot evolve with the changing surface density and the development of structures in the disc. The cooling approximation introduced by \citet{stamatellos2007} links the cooling rate directly to the local structure {\edit by assuming that each parcel of gas is embedded within a polytropic spherical pseudo-cloud}, providing a physically motivated approach for a modest additional computational cost. This method works well for spherical geometries, offering significant improvement on both $\beta$-cooling and barotropic models \citep{stamatellos2009thermo}, and was developed further by \citet{lombardi2015} {\edit with a different approach to estimating the depth of a gas parcel within the pseudo-cloud. This effectively improved the accuracy of the method for disc geometries \citep{mercer2018}. }

The same approach was recently employed in the ``modified Lombardi'' method developed by \citet{young2024} but now specifically tailored for self-gravitating discs. The modified Lombardi method involves estimating the column density above a parcel of gas from the pressure gradient and then adjusting for the reduction in scale height of a self-gravitating disc compared to a passive disc in which the scale height is set exclusively by stellar heating. This method produces excellent estimates of the column density throughout the disc and therefore can give accurate estimates of the optical depth {\edit in thick discs}, from which the radiative cooling rate (or heating rate) can be calculated. {\edit This cooling model can be coupled with radiation transport within the disc using the flux-limited diffusion (FLD) approximation in a ``hybrid'' method \citep{forgan2009}. The transfer of energy between neighbouring fluid elements in optically thick regions is therefore treated in addition to the radiative cooling that dominates where $\tau \sim 1$. Of course, a full radiative transfer calculation is necessary for complex structures that produce shadows, for example. The advantage of the hybrid radiative cooling approximation and FLD is the greatly reduced computational expense compared to full radiative transfer while maintaining accuracy for the disc structures we wish to study. }

The properties of the disc and star determine its stability to GI. Of particular interest is the range of disc-to-star mass ratios and radial extents for which a disc is susceptible to fragmentation or the development of spiral waves when realistic stellar heating is considered. By knowing this, we can infer whether direct fragmentation is a likely planet formation pathway, the influence of stellar heating on the disc's behaviour, and at what stages spiral waves might drive dust trapping, for example. In this paper, we re-examine the parameter space to investigate the properties of discs that develop GI using the more accurate approximate cooling treatment of \citet{young2024}.

The method, including the implementation of the radiative cooling approximation with four different approaches, is described in detail in section \ref{sec:method}.
We then test the polytropic cooling approximation methods introduced by \citet{young2024} and compare the outcome of simulations conducted with each of the methods. These are presented in section \ref{sec:comparemethods}.  
Next, we conduct a parameter study with the most accurate of those methods ("modified Lombardi") to explore whether and how the range of disc and stellar parameters for which the GI is active changes compared to prior work (section \ref{sec:parameterstudy}). The key results are discussed in detail with reference to observations and future modelling in section \ref{sec:discussion} and are summarised in section \ref{sec:conclusion}.

\section{Numerical method}
\label{sec:method}

{\edit The simulations were performed with a version of {\sc phantom} \citep{price2018aa} that has been modified to combine a radiative cooling approximation with radiative transfer using the flux-limited diffusion method \footnote{The code is available from \url{https://github.com/alisonkyoung1/phantom}.}. This hybrid approach was introduced by \citet{forgan2009} and here we implement FLD in tandem with improved versions of the radiative cooling approximation that were described in \citet{young2024}. }

\subsection{Radiative cooling approximation}
\label{sec:radcool}

The radiative cooling rate is approximated by estimating the optical depth throughout the disc. This approach was introduced by \citet{stamatellos2007}, extended to include flux-limited diffusion {\edit(FLD)} radiative transfer by \citet{forgan2009}, and further refined by \citet{lombardi2015}. A key limitation of this approach is the accuracy of the optical depth estimate, derived from either the local gravitational potential or the local pressure gradient. \citet{young2024} introduced two new approaches which give estimates of the local column density far closer to that calculated directly from the particle distribution. A more accurate value of column density gives a much better estimate of the optical depth, and therefore of the radiative cooling (or heating) rate. The four methods for estimating the pseudo-mean column density $\Bar{\Sigma_i}$ are described below.

\begin{enumerate}
    \item Stamatellos method. The column density is estimated from the local gravitational potential $\psi_i$ and density $\rho_i$ of particle $i$ \citep{stamatellos2007}  :
    \begin{equation}
    \label{eq:avcoldensity_stam}
    \Bar{\Sigma}_i = \zeta_n \left[ \frac{-\psi_i \rho_i}{4 \pi G}\right]^{1/2}.
\end{equation}
    $\zeta_n$ is a factor that depends on the polytropic index $n$ \citep[see][]{stamatellos2007}. Here we use $\zeta_2=0.368$. 
    
    \item Lombardi method. The local pressure $P_i$ and hydrodynamic acceleration, $ \boldsymbol{a}_{{\mathrm h},i} = - \nabla P_i /\rho_i$, which contains the pressure gradient, can be shown to give \citep{lombardi2015}:
    \begin{equation}
    \label{eq:coldens_lom}
       \Bar{\Sigma}_i = \frac{\zeta ' P_i }{\left | \boldsymbol{a}_{{\rm h},i} \right |},
    \end{equation}
    where the factor $\zeta '=1.014$
    
    \item Combined method. The pressure scale height is obtained 
    for the Stamatellos method of estimating $\Bar{\Sigma_i}$ (eq.~\ref{eq:avcoldensity_stam}),
    \begin{equation}
     H_{{\rm S}, i} = \Bar{\Sigma}_i /\zeta '  \rho_i .
     \end{equation}
    Then similarly, we obtain the scale height via the Lombardi method,
    \begin{equation}
      H_{{\rm L},i} = \frac{P_i}{\rho_i \left| \boldsymbol{a}_{{\mathrm h},i} \right|}.
\end{equation}
    These two values are then averaged in inverse quadrature:
    \begin{equation}
    H_{\rm C} = \left(H_{\rm P}^{-2} + H_{\rm S}^{-2}\right)^{-1/2}. \label{eq:combined}
\end{equation}
    This improves the estimates of scale height and the derived column density since the result tends to the lower of the two values. This means that for most of the disc, the Lombardi estimate dominates but where the pressure gradient approaches zero (near the mid-plane), the Stamatellos estimate takes over \citep{young2024}.

    \item Modified Lombardi: The value obtained with the Lombardi method is averaged in quadrature with the analytical value of scale height in a self-gravitating disc. In a self-gravitating disc the scale height $H_0$ is reduced compared to a non-self-gravitating disc in which $H_* = c_s/\Omega_{\rm K}$, where $c_s$ is the local sound speed. The ratio can be found to be \citep{young2024}

    \begin{equation}
    \frac{H_0}{H_*} = \frac{\sqrt{\pi/2}}{\sqrt{1 + 1/ \left( Q_{\rm 3D} \sqrt{\pi/2} \right)}}. \label{eq:disc_H_approx}
    \end{equation}

   Here,
    \begin{equation}
        Q_{3D} = \Omega_{\rm K}^2/4\pi G \rho(0),  
    \end{equation}
    is the 3D Toomre Q parameter \citep{mamatsashvili2010}, with the mid-plane density $\rho(0)$.
    
   The modified Lombardi scale height is calculated similarly to equation \ref{eq:combined} but replacing $H_{\rm S}$ with $H_0$.
\end{enumerate}

In section \ref{sec:comparemethods} we compare the evolution of a disc under the four methods. In the rest of the paper, we apply the modified Lombardi method since this provides the closest estimate of local column density to the actual value, both in the disc and within clumps.

{\edit Once we have estimated the column density, we can find the optical depth. 
The pseudo-mean optical depth is given by
\begin{equation}
    \label{eq:taumean}
    \bar{\tau} = \bar{\Sigma_i} \bar{\kappa_i}(\rho_i,T_i).
\end{equation}

\noindent A look-up table of the pseudo-mean mass opacity $\bar{\kappa_i}(\rho_i,T_i)$ is pre-calculated by averaging the Rosseland mean opacity over the pseudo-cloud; see \citet{stamatellos2007} and \citet{lombardi2015} for the details.}

{\edit  The radiative cooling rate is then:
\begin{equation}
     \label{eq:coolingrate}
    \frac{du_i}{dt}\Big| _{\rm rad} = \frac{4 \sigma_{\rm B} \left(T_{0}^4(\boldsymbol{r}_i) - T_i^4 \right)}{\Bar{\Sigma_i}^2\Bar{\kappa_i}(\rho_i,T_i) + \kappa_i^{-1}(\rho_i,T_i)},
\end{equation}
where $\sigma_{\rm B}$ is the Stefan-Boltzmann constant, $T_i$ is the temperature of the $i$th particle, and $\kappa_i^{-1}(\rho_i,T_i)$ is the Planck mean opacity. $T_0$ is the background temperature towards which the parcel of gas is cooling or heating. Here, this term is set by considering the stellar luminosity $L_*$ so as to incorporate the effect of stellar heating:
\begin{equation}
    T_{0,{\rm } i}^4 = T_{\rm floor}^4 + \exp(-\Sigma_i \bar{\kappa}_i ) \frac{L_*}{16 \pi \sigma_{\rm B} r_i^2}.
    \label{eq:discmintemp}
\end{equation}
\noindent The background heating due to the interstellar radiation field is assumed to give $T_{\rm floor} = 5$~K and the additional local stellar heating at radius $r$ is found using the stellar luminosity $L_*$, the local column density $\Sigma_i$ and the mean opacity $\bar{\kappa}_i$ \citep[see][]{stamatellos2007}. While the attenuation factor $\exp{(-\Sigma_i \bar{\kappa_i})}$ significantly reduces the heating affect of stellar irradiation near to the mid-plane, heat is transported from higher up by the diffusion term.}

{\edit The equilibrium temperature is determined by assuming
\begin{equation}
\label{eq:eqcondition}
  \frac{du_i}{dt}\Big| _{\rm hydro}  + \frac{du_i}{dt}\Big| _{\rm rad} + \frac{du_i}{dt}\Big| _{\rm FLD} = 0, 
\end{equation}
where ``hydro'' refers to the $pdV+$ viscous contribution.}

{\edit The equilibrium temperature is thus
\begin{equation}
    \label{eq:eqtemp}
    T_{\rm{eq},i}^4 = T_{0}^4(\boldsymbol{r}_i)+ \frac{1}{4\sigma}\left(\Bar{\Sigma}^2\Bar{\kappa_i}(\rho_i,T_i) + \kappa_i^{-1}(\rho_i,T_i)\right)
    \left[ \frac{du_i}{dt}\Big| _{\rm hydro} + \frac{du_i}{dt}\Big| _{\rm FLD}\right] .
\end{equation}
We can then set the equilibrium internal energy $u_{\rm eq,i} = u(T_{{\rm eq},i},\rho_i)$. A thermal timescale is defined:
\begin{equation}
    \label{ttherm}
    t_{{\rm therm},i} = \left(u_{\rm{eq},i} - u_i \right) \left [\frac{du_i}{dt}\Big| _{\rm hydro} + \frac{du_i}{dt}\Big| _{\rm rad} + \frac{du_i}{dt}\Big| _{\rm FLD}\right]^{-1}.
\end{equation}
The equilibrium internal energy and thermscale are used to evolve the energy equation by considering the progress of the energy of gas particle $i$ towards the equilibrium value. }

{\edit We implement the energy update slightly differently to \citet{young2024}, solving the update directly in the leapfrog integrator, for the active hydrodynamical sub-time-step $\delta t$. For the forward sub-time steps the energy is evolved with:
\begin{equation}
        u_{i+\frac{1}{2}} = u_i \exp(-\delta t/t_{\rm{therm},i})+ u_{\rm{eq},i}\left[1-\exp(-\delta t/t_{\rm{therm},i}) \right ].
\end{equation}
\noindent For the reverse sub-time steps:
\begin{equation}
    u_{i-\frac{1}{2}} = \frac{u_i - u_{\rm{eq},i} \left[ 1-\exp(-\delta t/t_{\rm{therm},i})\right]}{\exp(-\delta t/t_{\rm{therm},i}) }.
\end{equation}
}

\subsection{Simulations}
Simulations are performed with the smoothed particle hydrodynamics (SPH) code {\sc phantom} \citep{price2018aa} using the radiative cooling method implemented by \citet{young2024}, and detailed in section \ref{sec:radcool}. {\sc phantom} is an established astrophysical code that has been used extensively to model gravitationally unstable discs with the $\beta$-cooling method \citep[e.g.][]{rowther2023hiding}, the polytropic cooling approximation \citep[e.g.][]{cadman2020a}, and more recently with coupled ray-tracing radiative transfer \citep{rowther2024b}.

The central star is modelled as a sink particle \citep{bate1995aa} with the accretion radius set to 1~au, independently of the stellar mass. The luminosity of the host star is set according to the MIST stellar evolution models at 0.5~Myr \citep{choi2016,dotter2016} following \citet{haworth2020} (see Table \ref{tab:luminosity}). {\edit The heating due to irradiation from this central star is implemented via equation \ref{eq:discmintemp}.}

The artificial viscosity parameter $\alpha_{\rm{AV}}$ is varied between 0.01 and 1.0 using the switch of \citet{cullen2010}
and we set $\beta_{\rm{AV}} = 2.0$ to keep numerical dissipation low \citep{lodato2011} and to prevent over-reducing viscosity at shocks \citep{meru2012}.

\subsection{Initial Conditions}
\label{sec:initconds}
We aim for an initial structure that is consistent with that of a self-gravitating and radiatively cooled/heated disc. This is to avoid an initial disc structure that is overdense and artificially unstable to fragmentation. To this end, we implement an altered setup routine in {\sc phantom} to initialize the sound speed $c_{\rm s}$ using the luminosity of the host star. As usual, the sound speed profile is initialized with 
\begin{equation}
c_{\rm s}(r) = c_{\rm s,0} (r/r_0)^{-q}, 
\end{equation}
\noindent where $c_{\rm s,0}$ is the sound speed at $r=r_0$. The exponent $q=1/4$, following the analytical temperature profile for a passively irradiated disc. The sound speed at $r_0$ is

 \begin{equation}
     c_{\rm s,0} = \frac{1}{\sqrt{\pi/2}} \sqrt{\frac{k_{\rm B}}{\mu m_p}} \left[ \left(\frac{L_{*}}{4\pi\sigma_{\rm B} r_0^2} \right) + T_{\rm floor}^4 \right] ^\frac{1}{8}.
 \end{equation}

\noindent The prefactor $1/\sqrt{\pi/2}$ is from $H/H_* = \sqrt{\pi/2}$ where $H$ is the scale height of a self-gravitating disc and $H_* = c_{\rm s}/\Omega_{\rm K}$. This is found to produce sensible initial conditions (see Appendix \ref{sec:app_init}).

The initial surface density profile of the disc is set via

\begin{equation}
    \Sigma(r) = \Sigma_0 \left ( \frac{r}{r_0} \right ) ^{-p} \exp{ \left [ \left ( \frac{r}{r_c} \right ) ^{2-p}\right ]} \left (1-\sqrt{1/r} \right ),
\end{equation}

\noindent where $\Sigma_0$ is the surface density at $r=r_0$ and $r_c$ is the characteristic radius chosen to taper the density towards the outer edge. $\Sigma_0$ is set according to the chosen disc mass. We set $p=1.0$ and taper the density profile with $r_c = 0.9 r_{\rm{out}}$ such that the radial density profile is stable as the disc begins to evolve. Unless otherwise stated, simulations are initialised with $5 \times 10^5$ SPH particles. This resolution is sufficient to determine the large-scale evolution of the disc, while also keeping run times short enough to undertake a parameter study of over 60 simulations (see Appendix \ref{sec:resolution}).

\begin{table}
    \centering
      \caption{Stellar luminosity used for the simulations assuming an age of 0.5~Myr.}
    \label{tab:luminosity}
    \begin{tabular}{c|c}
        $M_*$ (M$_\odot$) & $L_*$ (L$_\odot$) \\
        \hline
         0.1 & 0.074 \\
         0.2 & 0.275 \\
         0.3 & 0.575 \\
         0.5 & 1.143 \\
         0.7 & 1.900 \\
         1.0 & 3.310 \\
    \end{tabular}
\end{table}

\begin{table}
	\centering
	\caption{Model parameters.}
	\label{tab:parameters}
	\begin{tabular}{cc} 
		\hline
		Parameter &  Value(s)\\
		\hline
		$M_{*}$ $\left(\rm{M}_{\odot}\right)$ & 0.1, 0.2, 0.3, 0.5, 0.7, 1.0 \\
		$M_{\rm d}/M_{*}$ & \{0.1:1.4\} \\
		$r_{\rm{out}}$ (au) & 50, 100, 200 \\
        $p$ & 1  \\
        $q$ & $1/4$ \\
        $T_{\rm floor}$ (K) & 5 \\
	\end{tabular}
\end{table}

\subsection{Passive disc test}
\begin{figure}
    \centering
    \includegraphics[width=1\linewidth]{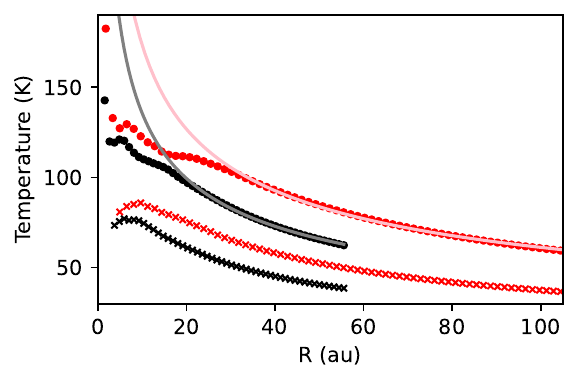}
    \caption{The radial temperature profiles for two simulated low-mass (i.e. largely passively irradiated) discs showing the {\edit mass-weighted} vertically 1averaged temperature (circles), fitted profiles (solid lines), and mid-plane temperature (crosses). Black: star of \solmass{0.5}, disc of \solmass{0.005}, L=\sollum{1.143}, T$_{\rm floor}=$~5~K, and initial radius 50~au. Red: star of \solmass{1}, disc of \solmass{0.01}, L=\sollum{3.31}, T$_{\rm floor}=$~5~K, and initial radius 100~au. The fitted lines are $337  r^{-0.41}$ (black) and $491 r^{-0.45}$. }
    \label{fig:passiveTvsR}
\end{figure}

First, we verify that temperatures are accurately estimated in the model by testing the method on passive discs, i.e. discs in which the temperature is set by the incident stellar radiation and the internal heating is negligible. Fig.~\ref{fig:passiveTvsR} shows the radial profiles of the vertically averaged temperature for two simulated discs with $M_{\rm disc}/M_{*}=0.01$. The simulation data was fitted with the profile $T=T_0 r^{-b}$, where $T_0$ is the temperature at $r=1$~au, the disc inner edge. The temperature profiles are well described by power laws with $b=0.41$ and $b=0.45$. These values are close to the analytical value expected for passively irradiated discs ($b\approx 0.43$, \citealt{chiang1997}). The crosses show the mid-plane temperature for each disc, found by averaging only particles with $|z| < 0.5 H$. The temperature at the mid-plane is substantially reduced by the attenuation term in equation \ref{eq:discmintemp}. Temperatures are underestimated in the inner regions of the disc but we do not expect this to affect the results. {\edit In section \ref{sec:comparemethods} we compare the mid-plane temperatures with those derived from observations.}

\section{Results}
\label{sec:results}

First, we present the results of comparing simulations with the four different approaches to estimating the optical depth in section \ref{sec:comparemethods}. Following this, the outcomes of the parameter study using the modified Lombardi are presented in section \ref{sec:parameterstudy}. There, we describe the differences in disc evolution when simulated with the improved method.

In this analysis we estimate \(\beta_{\rm{cool}} = \Omega_{\rm K} t_{\rm{cool}}\) by using an estimate of the cooling time in annuli of the disc. In these simulations, heat is transferred by both flux-limited diffusion and radiative cooling via the polytropic cooling approximation. {\edit The cooling timescale of a disc annulus at $r$ is therefore given by:
\begin{equation}
    t_{\rm{cool}} (r)= \sum_i \frac{-(u_i-u_{0,i})}{\dot{u}_{{\rm rad},i} + \dot{u}_{{\rm FLD},i}}.
    \label{eq:tcool}
\end{equation}
\noindent where $u_{i}$ is the internal energy density of gas particle $i$, $u_0 = u(T_{0,i},\rho_i)$ and $\dot{u}_{{\rm rad},i}$ and $\dot{u}_{{\rm FLD},i}$ are the heating rates due to radiative heating/cooling and FLD respectively. Only particles with $|z|< 0.5H$ are included in the summation so we can probe cooling times near to the mid-plane.} To allow comparison with the results of simulations that employed a variable $\beta_{\rm{cool}}$ of the form 
\begin{equation}
    \beta_{\rm{cool}}(r)= a (r/r_{\rm{in}})^{-b},
\end{equation}

\noindent we fit the values calculated from our simulations, $\log(|\beta_{\rm{cool}}|)$, with the function 
\begin{equation}
\label{eq:betafunc}
   f(r,a,b) = \log(a) -b\log(r/r_{\rm in}).
\end{equation}

The Shakura-Sunyaev viscosity $\alpha_{\rm ss}$ provides a parameterized form of viscosity resulting from unresolved turbulence within a disc \citep{shakura1973}. Gravitational stresses contribute to transport angular momentum transport in self-gravitating discs \citep{lynden-bell1972} and it is useful to compare the resulting equivalent values of $\alpha_{\rm ss}$:
\begin{equation}
 \alpha_{\rm grav}(r) = 
    \left(\frac{d \ln{\Omega_{\rm K}}}{d \ln{r}} \right)^{-1} \frac{T^{\rm grav}_{r,\phi}}{\Sigma c_s^2} 
    \end{equation}
For each annulus in the disc, the gravitational stress $\alpha_{\rm grav}$ is estimated by summing the radial and azimuthal components of the gravitational acceleration $g_r$ and $g_\phi$ over $N$ particles in the annulus, which gives:
\begin{equation}
   \alpha_{\rm grav}(r) =\left(\frac{d \ln{\Omega_{\rm K}}}{d \ln{r}} \right)^{-1} \frac{1}{4\pi G \langle c_{\rm s}\rangle^2}   \frac{1}{N}\sum^N_{i=1} \frac{g_{r,i} g_{\phi,i}}{\rho_i}.
    \label{eq:alpha_grav}
\end{equation}

\begin{figure*}
    \centering
    \includegraphics[trim=3.8cm 7cm 4.7cm 1.5cm,clip, width=0.9\linewidth]{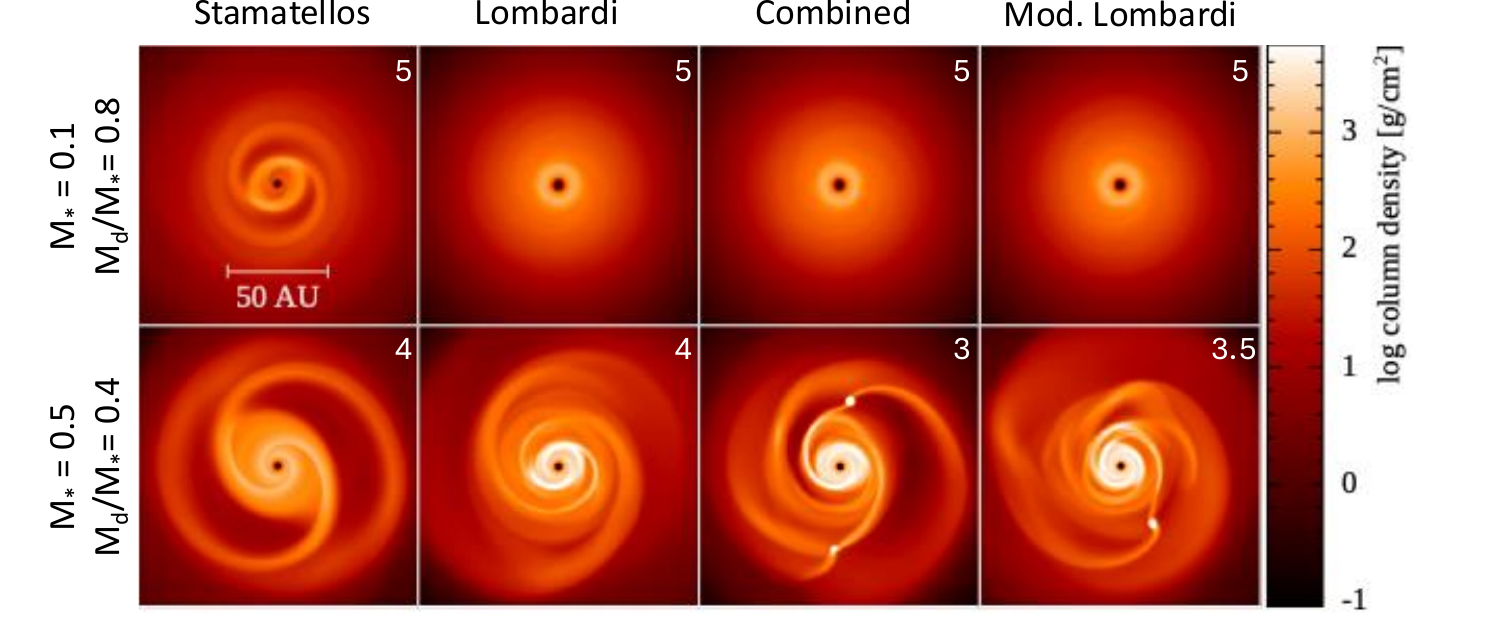}
    \includegraphics[trim=0cm 3.2cm 0cm 2.8cm,clip, width=0.9\linewidth]{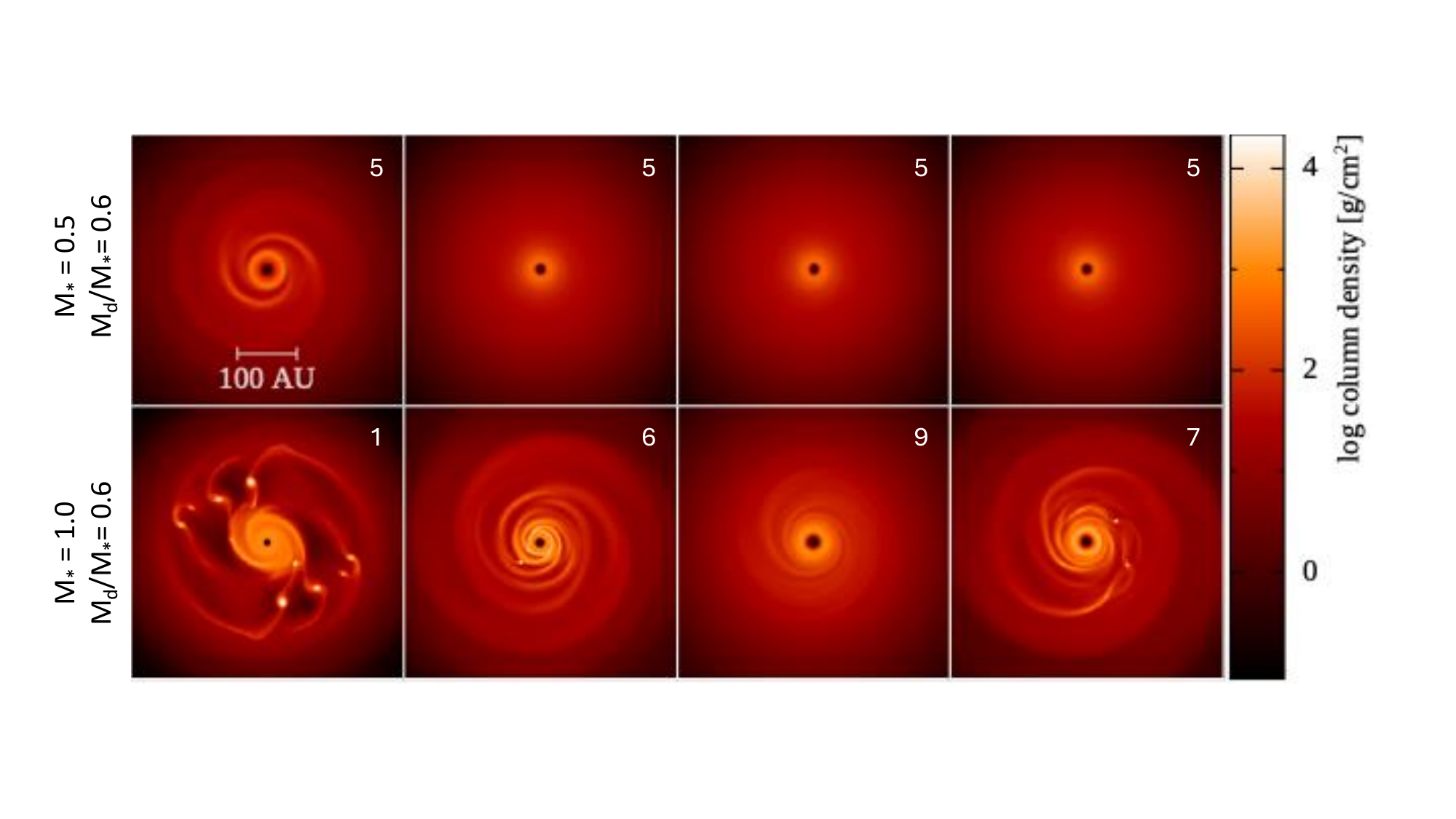}
    \caption{Snapshots from simulations performed with the four methods for four sets of initial conditions with $R_{\rm out}=50$~au (upper two rows) and $R_{\rm out}=200$~au (lower two rows). The time stamp of each snapshot is given in terms of outer rotation periods for each snapshot in white. If fragmentation occurred, the simulation was stopped shortly afterwards because the simulations run very slowly once clumps form. For the more stable disc-to-star mass ratios, the methods produce similar results but discs are generally more susceptible to gravitational instability with the Stamatellos method. More accurate treatment of radiative cooling is more important for discs that are closer to instability.}
    \label{fig:CompMethodsDens}
\end{figure*}

\subsection{Comparison of radiative cooling approximation methods}
\label{sec:comparemethods}

 In section \ref{sec:radcool}, we described four approaches for a radiative cooling approximation. We now present the outcomes of simulations employing each method for comparison. Fig.~\ref{fig:CompMethodsDens} shows column density renderings of discs evolved using the four methods from four sets of initial conditions. In the first and third rows, the Stamatellos method leads to gravitational instability, evident as spiral waves, while the other methods lead to a stable, axisymmetric disc. The simulations on the second and fourth rows display other differences in evolution. For the discs initialised with $M_*=$~\solmass{1}, $R_{\rm{out}}=200$~au, and $M_{\rm{d}}/M_* = 0.6$ (Fig.~\ref{fig:CompMethodsDens}, fourth row), the disc quickly fragments under the Stamatellos method but remains stable under the Lombardi and combined methods. With the modified Lombardi method the disc maintains a stable spiral structure but forms a fragment after seven outer rotation periods. We now examine this latter model further in an effort to explain the differences in evolution.

\begin{figure}
    \includegraphics[width=0.95\columnwidth]{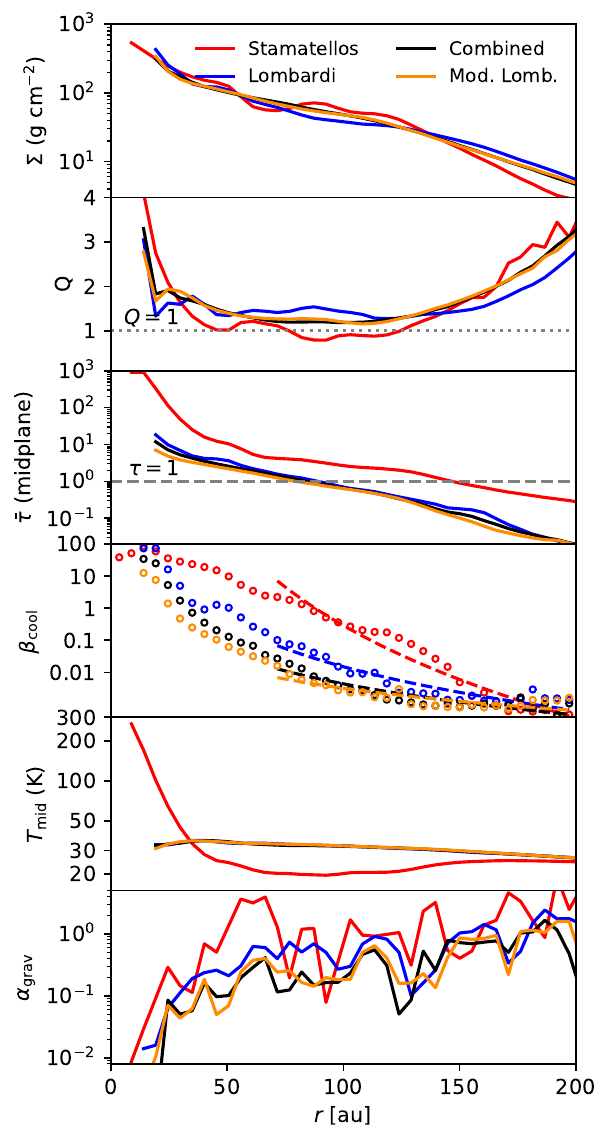}  
    \caption{Azimuthally averaged values of the surface density $\Sigma$, Toomre $Q$ parameter, mid-plane pseudo-mean optical depth ($\bar{\tau}$), and $\beta_{\rm{cool}}$, mid-plane temperature $T_{\rm mid}$, and gravitational stress $\alpha_{\rm {grav}}$ for the four methods for the simulations with \solmass{1} star, $R_{\rm{out}}=200$~au, and $M_{\rm{d}}/M_* = 0.6$ (bottom row of Fig.~\ref{fig:CompMethodsDens}). The graphs were constructed by averaging simulation outputs taken before clumps formed in the Stamatellos simulation and after 5 ORPs for the others. 
    The azimuthal averages of $\tau$, $\beta_{\rm{cool}}$, and $T_{\rm mid}$ were calculated using only particles with $|z| < H/2$. For the $\beta_{\rm{cool}}$ panel, the dashed lines show the fitted functions of the form of equation \ref{eq:betafunc} to $\beta_{\rm{cool}}$. The fits give $\beta_{\rm{cool}} \propto r^{-x}$, with $x\approx$ 9, 5, 3, 2 respectively for the Stamatellos, Lombardi, Combined and modified Lombardi methods. }
    \label{fig:betaplot_compare}
\end{figure}

The differences in the evolution are caused by the differences in estimates of optical depth, which affects both the minimum mid-plane temperature (c.f. equation \ref{eq:discmintemp}) and how easily heat generated by the weak spiral shocks can be radiated away. {\edit The pseudo-mean optical depth $\bar{\tau}$ (Eq. \ref{eq:taumean}), $Q$, effective $\beta_{\rm cool}$, mid-plane temperature  and gravitational stress $\alpha_{\rm grav}$ for the simulations with \solmass{1} star, $R_{\rm{out}}=200$~au, and $M_{\rm{d}}/M_* = 0.6$ (bottom row of Fig.~\ref{fig:CompMethodsDens}) are presented in Fig.~\ref{fig:betaplot_compare}. }  To study the radial profiles, these quantities are estimated in radial bins using equations \ref{eq:toomreQ}, \ref{eq:tcool} and \ref{eq:alpha_grav} respectively.

{\edit Disc mid-plane temperatures are difficult to infer from observations because of the high optical depths. Nevertheless, estimates for Class 0/I discs range from $\sim$20-70~K within $\sim 50$ au \citep[e.g.][]{van-t-hoff2018ab,vanthoff2020,zamponi2021,takakuwa2024,maureira2025}. We note that the disc mid-planes in our simulations with all methods are warm at $\approx$~20-30~K, which is in agreement with observations of young discs.}

The Stamatellos method gives rise to a largely optically thick disc, with the mid-plane optical depth dropping below $\bar{\tau}=1$ at $R=150$~au compared to within 100~au for the other methods. 
This is despite the discs having a similar surface density profile and the reasons for the difference, discussed previously in \citet{young2024} and \citet{mercer2018}, are to do with the the assumption of spherical symmetry in estimating the column density from the gravitational potential which results in an overestimate. The mid-plane temperature is correspondingly lower, due to the additional attenuation of the stellar irradiation.
For all four simulations, $\beta_{\rm cool}$ is low enough to allow fragmentation where $Q \lesssim1$. The cooler mid-plane obtained with the Stamatellos method allows $Q$ to decrease such that the disc becomes unstable and fragments. In the other simulations, the temperature is just high enough for a quasi-stable spiral structure to be maintained. After several outer rotation periods, localised regions become unstable and collapse in the modified Lombardi simulation. In all four simulations, $\alpha_{\rm grav}> 0.1$ in most of the disc, showing that gravitational torques are driving the disc evolution, and $\alpha_{\rm grav}$ is noteably higher for the Stamatellos simulation.

We estimate the radial exponent $b$ of $\beta_{\rm{cool}}$ by fitting equation \ref{eq:betafunc} between 70 and 200 au and this is shown in the fourth panel of Fig.~\ref{fig:betaplot_compare}. {\edit $\beta_{\rm{cool}}$ decreases most steeply with the Stamatellos method and for the other methods it flattens off towards in the outer disc.}

For more stable discs, we find the evolution of the Lombardi, combined and modified Lombardi methods to be similar, but this often differs from the evolution under the Stamatellos method. In situations where the system is close to fragmentation, there may be small differences between the former three methods in whether fragments form, time to fragmentation, or number of fragments.

\subsection{Fragmentation and stability in discs around low mass stars}
\label{sec:parameterstudy}

\begin{figure}
    \centering
    \includegraphics[width=0.9\columnwidth]{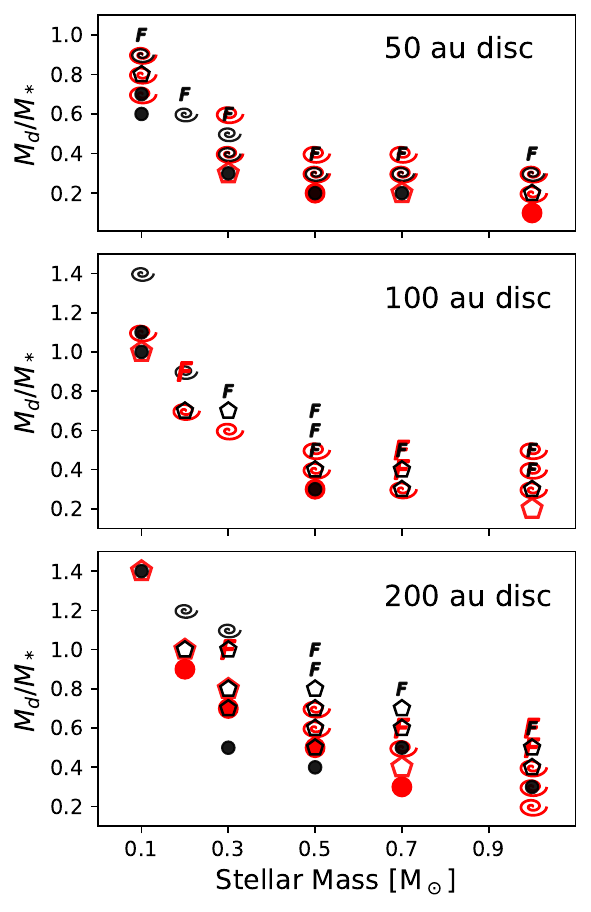}
    \caption{The outcomes of the grid of simulations, illustrating whether the disc remained axisymmetric (circles), developed a faint spiral (pentagons), developed an extended spiral (icons), or fragmented (F). The discs which fragmented developed strong spiral structures as well. The results with the modified Lombardi method are shown in black and the with the Stamatellos method are plotted in red.}
    \label{fig:gridoutcomes}
\end{figure}

Following \citet{haworth2020} and \citet{cadman2020a}, we simulated discs around stars of masses \solmass{0.25} to \solmass{1} including heating from the central star. We employ the same values of stellar luminosity that they estimated from the MIST stellar evolution tracks at 0.5~Myr (\citealt{choi2016,dotter2016}, Table \ref{tab:luminosity}). Like \citet{cadman2020a} and \citet{haworth2020}, we simulate discs with initial outer radii of 50, 100, and 200~au. {\edit The full initial parameters are given in Table~\ref{tab:parameters} and section \ref{sec:initconds}}. We initially ran the simulations for 5 outer rotation periods (ORP) but found that the structure evolved further after this time in some cases. For example, the \solmass{0.2}, 50~au disc around a \solmass{1.0} star remained axisymmetric for nearly 9 ORPs before a spiral developed. Interestingly, when employing the cooling of \citet{stamatellos2007} to compare with \citet{haworth2020}, we found this disc began to develop a spiral structure after 4 ORPs, which suggests \citet{haworth2020} may have observed the spiral form if the simulations had run for longer than a minimum of 3 ORPs. Consequently, we evolve the discs longer to at least 10 ORP to verify that the structure is steady. These differences could also be due to the initial temperature profiles which were set in the optically thin limit in the simulations of \citet{haworth2020} and \citet{cadman2020a}. For reference, the results of \citet{haworth2020} and our simulations with the Stamatellos method are compared in appendix \ref{sec:app_haworth}.

\begin{figure*}
    \centering
    \includegraphics[width=0.9\linewidth]{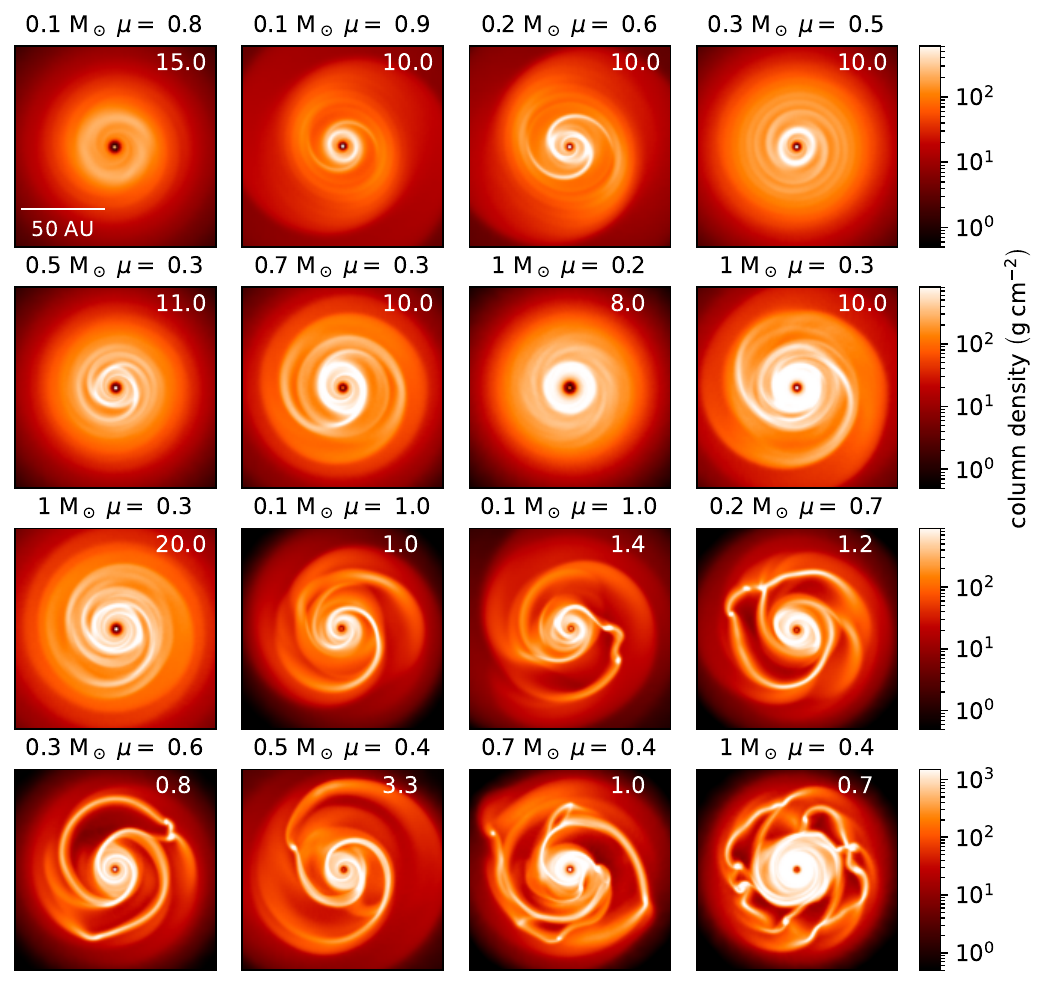}
    \caption{The morphology of spiral structures in discs with initial radius of 50 au, from simulations employing the modified Lombardi method. The snapshots displayed have been selected to show the range of structures that develop in the discs; the timestamp is given in the upper right of each panel in units of ORP. Stellar masses are given above each panel and $\mu = M_{\rm{d}}/M_* $. The structures in the \solmass{0.1} $\mu = 0.8$ and \solmass{1} $\mu=0.2$ simulations are classed as faint spirals.}
    \label{fig:R50spirals}
\end{figure*}

A summary of the outcomes of the simulations with the modified Lombardi and Stamatellos methods is presented in Fig.~\ref{fig:gridoutcomes}. We have divided the discs that formed spiral structures into two categories. Discs with ``faint spirals'' feature low contrast spiral arms, often with $m>2$ and more prominent in the inner disc regions. Discs with ``extended spirals'' have clear spiral arms throughout the entire disc. An examples of a faint spiral can be seen in the first panel of Fig.~\ref{fig:R50spirals} and extended spirals can be seen in the third and fourth panels of Fig.~\ref{fig:R50spirals}.

In general, we find that with the modified Lombardi method discs remain stable for a higher $M_{\rm{d}}/M_*$ than that found in prior work. As shown previously, lower mass stars can support discs of higher $M_{\rm{d}}/M_*$ than higher mass stars without the gravitational instability developing. However, the parameter space in which discs may fragment differs from earlier findings, and we describe this next.

\subsubsection{Fragmentation}

With the new modified Lombardi method, we observe fragmentation in relatively compact 50~au discs, even for a host stellar mass of \solmass{0.1}, in contrast to the Stamatellos method, in which 50~au discs were stable against fragmentation for the same parameters. We find that 50~au discs may be susceptible to fragmentation for $M_{\rm{d}}/M_* \gtrsim 0.4$. The more extended (and therefore less optically thick) discs also fragmented but only if $M_{\rm{d}}/M_* \gtrsim 0.5$. Fragments were produced between 20-30~au in the smallest discs which in our solar system is in the region of the orbits of Uranus and Neptune. If we consider that protoplanets will undergo inward migration, this allows the possibility of forming a wider variety of planets via direct fragmentation rather than just those on very wide ($>100$~au) orbits.

Some simulations like $M_*=$~\solmass{0.5} $R=50$~au $M_{\rm{d}}/M_* =0.4$ form just one collapsing clump but others quickly form ten or more. GI driven fragmentation can be expected to produce a range of outcomes: few or multiple collapsing clumps with semi-major axes of between $\sim20$ to $\sim 100$~au.

\subsubsection{Stability}
These simulations indicate that extended $R_{\rm{out}}=200$~au discs may be stable at higher masses than previously thought. The larger ($R_{\rm{out}}=$ 100 and 200~au) discs are strongly stabilised by stellar heating such that stars of $<$~\solmass{0.5} can support discs of the same mass as the star without significant development of gravitational instability. \citet{haworth2020} reported that a \solmass{0.4} disc around a \solmass{1} star will fragment, whereas we find that a disc as massive as \solmass{0.5} remains stable. For $M_*<$ \solmass{0.5}, 200~au discs do not fragment even if $M_{\rm{d}}/M_* > 1$. This indicates that low mass stars can support relatively massive extended discs. $M_{\rm{d}}/M_* \sim 0.1$ is typically quoted as indicating the conditions when the gravitational instability is likely to develop. However, these simulations indicate that spiral structures don't develop unless $M_{\rm{d}}/M_*$ is at least 0.4, and in most cases, much higher still.

\subsubsection{Spirals}

\begin{figure*}
    \centering
    \includegraphics[width=0.9\linewidth]{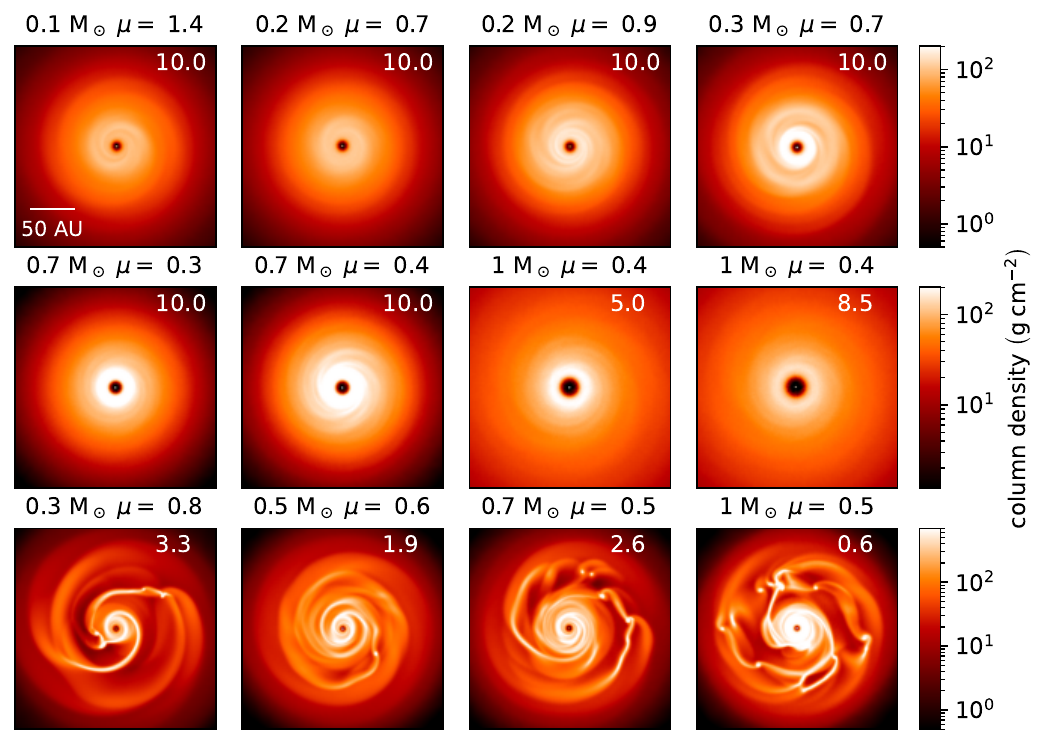}
    \caption{The morphology of spiral structures in discs with initial radius of 100 au. The timestamp is given in the upper right of each panel in units of ORP and $\mu = M_{\rm{d}}/M_* $.}
    \label{fig:R100spirals}
\end{figure*}

\begin{figure*}
    \centering
    \includegraphics[width=0.9\linewidth]{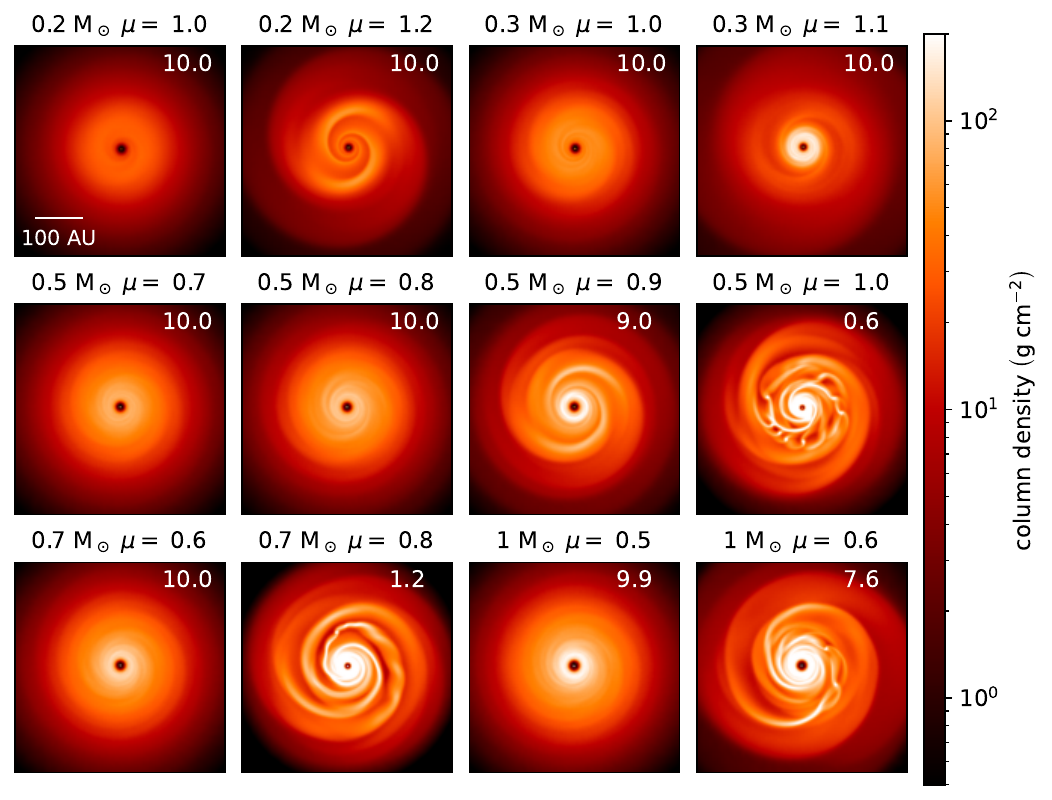}
    \caption{The morphology of spiral structures in discs with initial outer radius of 200 au. The timestamp is given in the upper right of each panel in units of ORP and $\mu = M_{\rm{d}}/M_* $.}
    \label{fig:R200spirals}
\end{figure*}

The variety of structures formed in the simulation can be seen in Figs.~\ref{fig:R50spirals},~\ref{fig:R100spirals} and \ref{fig:R200spirals}. In this paper, we classify the spiral structures as faint or extended. Nearly all the simulation snapshots in Fig.~\ref{fig:R50spirals} show extended spirals and the first two panels of Fig.~\ref{fig:R100spirals} are examples of faint spirals. Extended spiral structures form in a far narrower range of discs than found by \citet{haworth2020} or \citet{cadman2020a} because higher mass discs remain axisymmetric and fragmentation occurred rapidly in many cases, leaving few discs with sustained spiral structures. Faint spirals developed in some of the 100 and 200~au discs but the large-scale spirals typically associated with GI only formed for the lowest mass stars of $M_*\lesssim $ \solmass{0.5}.

\begin{figure}
    \centering
    \includegraphics[width=\columnwidth]{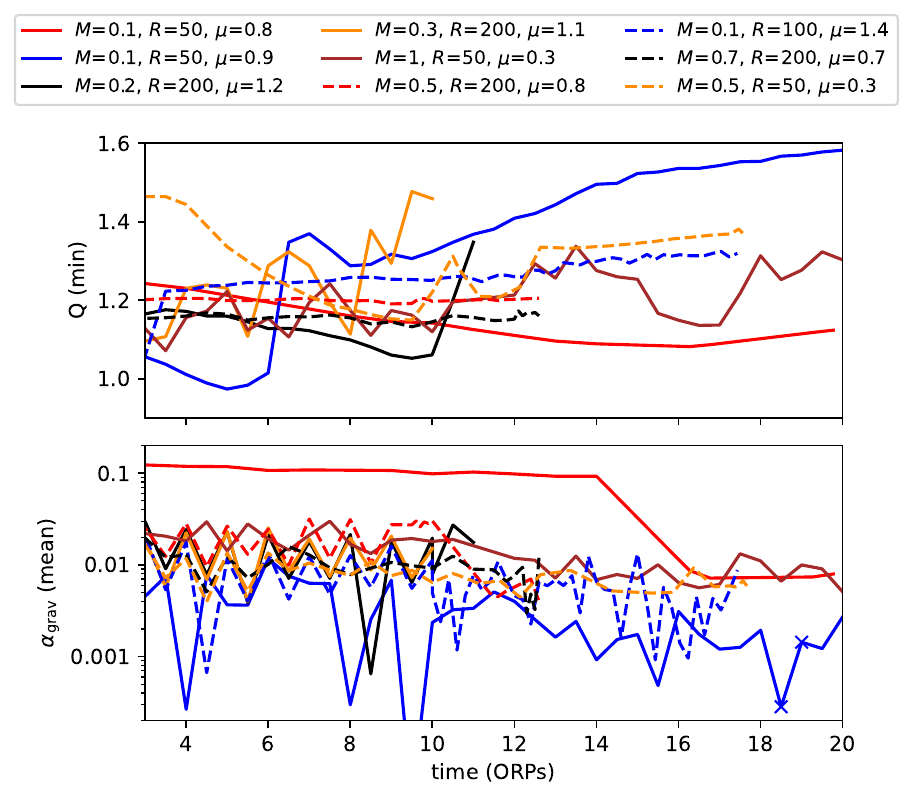}
    \caption{Long lived spiral structures. The time evolution of the azimuthally-averaged minimum value of $Q$ and the average value of $\alpha_{\rm grav}$ within $0.5R_{\rm out} <R< R_{\rm out}$. Solid lines denote discs with large scale, prominent spirals and dashed lines denote discs with faint spirals. Stellar masses are given in \solmass{}, outer radii in au, and $\mu = M_{\rm{d}}/M_* $. The points marked with `x' are negative values of $\alpha_{\rm grav}$, i.e. where the stresses drive outward net gas motions rather than accretion. The zig-zag pattern in $\alpha_{\rm grav}(\rm mean)$ is indicative of self-regulation.}
    \label{fig:Qminvt}
\end{figure}

Long lived spiral structures (at least 10 ORPs) form in discs of all three initial radii tested but are uncommon within the parameter space as a whole. The time evolution of the minimum values of $Q$ and the maximum values of $\alpha_{\rm grav}$ are plotted in Fig.~\ref{fig:Qminvt} for simulations with sustained spirals that were modelled for at least 10 ORPs. The mean gravitational stress is not constant but fluctuates on a dynamical timescale. This is evidence of self-regulation at work: Gravitational stresses heat the disc, increasing $Q$, weakening the instability, and reducing the stresses. The disc then cools back towards $Q = 1$, the instability strengthens, and the stresses increase \citep{clarke2009}. There may also be some non-local effects if these are high disc-to-star mass ratios, as discussed in \citet{lodatorice2005} and \citet{forgan2009}.
Some discs simulated had $Q_{\rm min} \sim 1$ for many ORPs but did not show the zig-zag stress pattern. Those discs are probably not self-regulating but are cooling towards either fragmentation or self-regulation.

\begin{figure}
    \centering
    \includegraphics[width=\columnwidth]{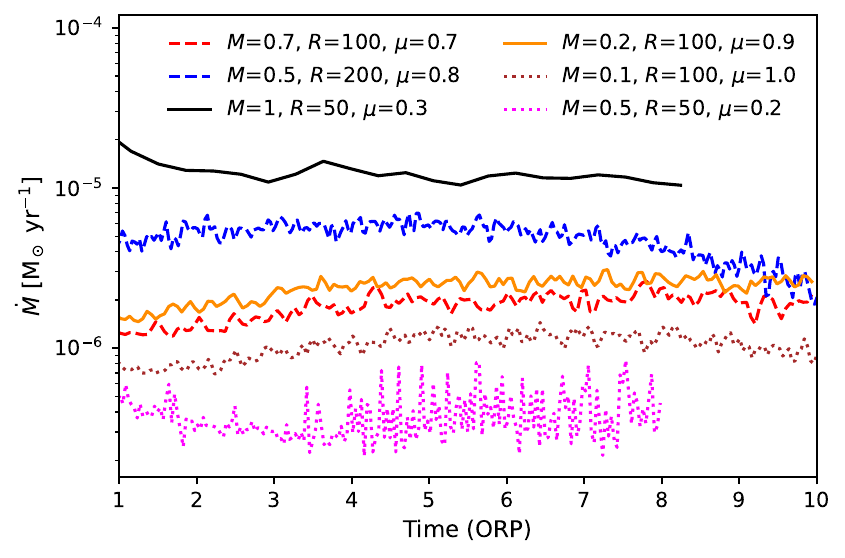}
   \caption{Accretion rates of selected discs smoothed by resampling every 20 years. The models were chosen to represent the range in accretion rates. Solid lines indicate discs which develop strong spirals, dashed lines indicate faint spirals, and dotted lines indicate discs that remain axisymmetric.}
    \label{fig:accretionrates}
\end{figure}

Examples of accretion rates over 10 ORPs are plotted in Fig.~\ref{fig:accretionrates}. The models were selected to illustrate the range of accretion rates of axisymmetric discs, discs with strong spirals and discs with faint spirals.
Discs with faint and strong spirals had higher accretion rates than axisymmetric discs, as expected, because of the additional effective viscosity due to the gravitational stresses. Discs with faint spirals appear to have accretion rates within a similar range to those with strong spirals, though we note that the disc masses differ between simulations. While accretion rates tend to be overestimated in SPH simulations, we do expect quasi-steady self-gravitating discs to have high accretion rates, probably of at least \solmass{$10^{-7}$}~yr$^{-1}$.


\subsubsection{Linking disc evolution to internal structure}
\begin{figure*}
    \centering
    \includegraphics[width=0.9\textwidth]{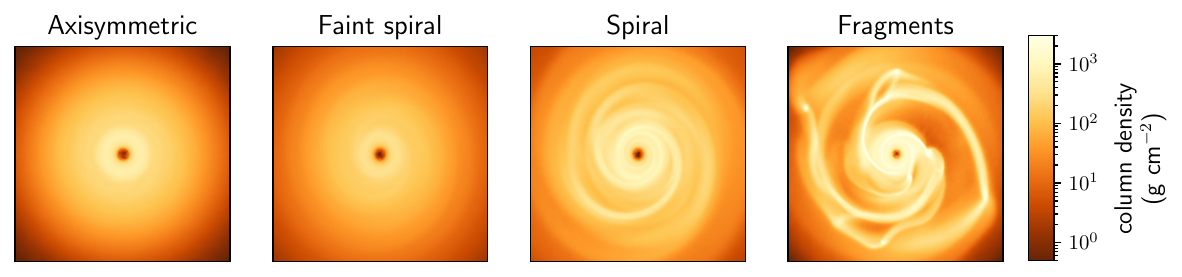}
    \caption{Examples of simulations displaying different morphology, for which various quantities are plotted in Fig.~\ref{fig:spatial_betaR50}. These simulations all had initial outer radii of 50~au and the width of each panel is 120~au. The simulations are from the left for: $M_*=$ \solmass{0.7}, $M_{\rm{d}}/M_* =0.3$;
     $M_*=$ \solmass{0.1}, $M_{\rm{d}}/M_* =0.8$;
     $M_*=$ \solmass{0.7}, $M_{\rm{d}}/M_* =0.3$;
     and $M_*=$ \solmass{0.7}, $M_{\rm{d}}/M_* =0.4$.}
     \label{fig:morphologysample}
\end{figure*}

\begin{figure}
    \centering
    \includegraphics[width=0.95\columnwidth]{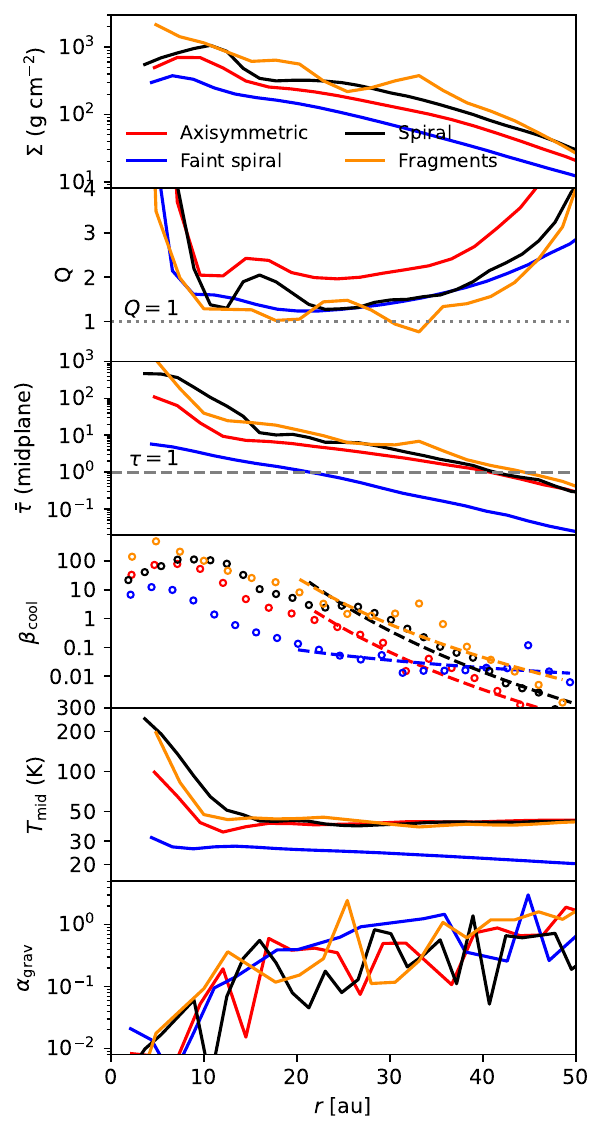}
    \caption{Azimuthally averaged values of the surface density $\Sigma$, Toomre $Q$ parameter, mid-plane pseudo-mean optical depth ($\bar{\tau}$), $\beta_{\rm{cool}}$, and mid-plane temperature $T_{\rm mid}$ for the $R=50$~au discs shown in Fig.~\ref{fig:morphologysample}. The red lines (axisymmetric) are for $M_*=$ \solmass{0.7}, $M_{\rm{d}}/M_* =0.2$; the blue lines (faint spiral) are $M_*=$ \solmass{0.1}, $M_{\rm{d}}/M_* =0.8$; the black lines (spiral) show $M_*=$ \solmass{0.7}, $M_{\rm{d}}/M_* =0.3$; and the yellow lines (fragments) show $M_*=$ \solmass{0.7},$M_{\rm{d}}/M_* =0.4$. 
    The values were extracted for time steps when structure has developed.}
    \label{fig:Q_tcool}
\end{figure}

We now compare the physical properties of discs that develop different morphology. The density structures of the four simulated discs that we will examine are shown in Fig.~\ref{fig:morphologysample} and azimuthally averaged physical properties are plotted in Fig.~\ref{fig:Q_tcool}.

For the axisymmetric disc, $Q>2$ and so there are no signs of GI-driven structure. The mid-plane optical depth is lowest in the disc with the faint spiral structure. Consequently, stellar heating is significant at the mid-plane for this disc. The disc is heated by the star to above the equilibrium temperature needed to support it against collapse and the disc has a near-isothermal temperature profile.

For discs with a higher column density, the mid-plane temperature is more sensitive to internal heating since the stellar heating is reduced there. In the discs with extended spiral arms, the disc mid-plane is optically thick within $R<40$~au. Radiative cooling and heating is efficient outwith these regions, and the equivalent $\beta_{\rm cool}\ll 1$, where the disc becomes optically thin (Fig.~\ref{fig:Q_tcool}).

\begin{figure}
    \centering
    \includegraphics[width=1\linewidth]{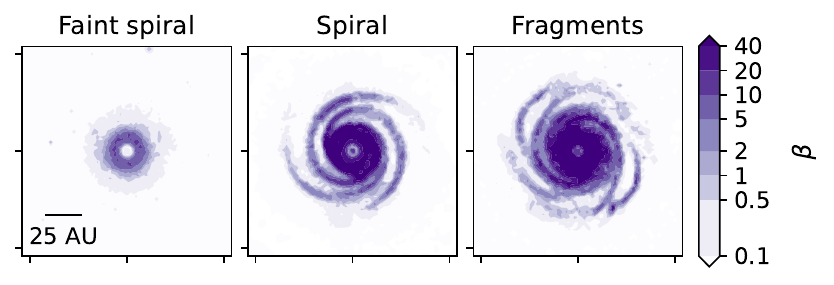}
    \caption{Estimated values of $\beta_{\rm cool}$ at the mid-plane for three of the simulated discs from Figs. \ref{fig:morphologysample} and \ref{fig:Q_tcool}. The left and centre panels show snapshots after 5 ORPs and the right-hand panel shows a snapshot shortly before fragmentation occurs. $\beta_{\rm cool}$ is sufficiently low in the dense spiral arms for them to collapse. }
    \label{fig:spatial_betaR50}
\end{figure}

Like in section \ref{sec:comparemethods}, we estimate the fit $\beta_{\rm{cool}} \propto R^{-x}$ between 20-50 au to sample the outer disc and these fits are shown as dashed lines in the fourth panel of Fig.~\ref{fig:Q_tcool}. The exponents obtained are approximately {\edit 10, 2, 11 and 9} for the axisymmetric, faint spiral, spiral and fragmenting simulations respectively. Estimates of $\beta_{\rm{cool}} $ in the disc mid-plane are also presented, in Fig.~\ref{fig:spatial_betaR50}, to show the spatial variation. $\beta_{\rm{cool}} $ is high enough towards the edge of the disc with the faint spiral to prevent large-scale spiral arms forming ($\beta_{\rm{cool}} \gtrsim20$).

\section{Discussion}
\label{sec:discussion}

We began by repeating simulations of protoplanetary discs with a variety of initial conditions with four different approximate radiative cooling methods. The evolution of self-gravitating discs is very sensitive to the radiative cooling (and heating) rate so it is important to estimate the optical depth accurately. This became evident in the outcomes of the simulations, in which discs remained axisymmetric when evolved with the more accurate methods compared to developing spiral structures in the original (Stamatellos) method because the latter method overestimates the optical depth. Not only are spiral structures less common with the more accurate methods but the nature of those spirals appears to be different. Most of the spirals shown in the results of \citet{haworth2020}, who also employed the Stamatellos method, are also large scale, low-$m$ spirals but in our wider exploration of the parameter space we find that these structures are uncommon (see Figs.~\ref{fig:R50spirals}, \ref{fig:R100spirals} and \ref{fig:R200spirals}). This points to a broader conclusion that the more accurate cooling methods tend not to produce grand design-like spirals, consistent with \citet{rowther2024b} and \citet{xu2025}.

\subsection{Discs around low mass stars can fragment}
It has been clear for some time that stellar irradiation impacts the evolution of massive protostellar discs but we have now studied a broad parameter space with an accurate approximate method to gain a deeper understanding of the implications. Our key finding is that stellar irradiation does not prevent discs from undergoing fragmentation or from developing self-regulating spiral structures, in agreement with several prior studies \citep{rice2011,mercer2018,mercer2020,cadman2020a,haworth2020,leedham2025}.

Within the parameters studied ($0.1<M_*<$\solmass{1.0}) fragmentation occurs more readily in discs around higher mass stars, as found by \citet{cadman2020a}. {\edit Analytical work has indicated that fragmentation is unlikely in discs of $R_{\rm out}=50$~au. For example, \citet{rafikov2005} found that disc conditions are only favourable for fragmentation beyond $\sim 100$~au; and \citet{clarke2009} found that fragmentation couldn't occur within $\sim 70$~au (their Figs. 5 and 6), unless the accretion rate is extremely high, at least 2 orders of magnitude higher than $\dot{m}$ measured in our simulations (Fig.~\ref{fig:accretionrates}).} We find that fragmentation is not restricted to $R>70$~au but can occur even in 50~au discs, with fragmentation occurring between $\sim30$-40~au \citep[see also][]{stamatellos2009,mercer2020}.



\subsection{Morphology}

When stellar heating strongly influences the disc temperatures, discs tend to remain axisymmetric or else fragment at higher $m_{\rm{disc}}/m_{\rm{star}}$. For lower stellar masses with lower luminosity ($m_* < 0.5 \rm M_\odot$ here), the disc evolution is less affected by stellar irradiation. In this case, spiral structures may form and discs remain stable to very high masses. The work of \citet{laughlin1996} predicted that the GI would predominantly drive the $m=2$ spiral mode, however we find that low-$m$ grand-design-type spirals appear to be less common. For discs in which stellar irradiation plays a greater role in setting the temperature, we see fainter, higher-order spirals. Elucidation of this effect will require higher resolution simulations of those discs, which we leave for future work.

\subsection{Long-lived spirals}

There has been some disagreement over whether a self-regulated thermal state exists. \citet{kratter2011} showed that discs around 1-3 Myr stars are irradiation dominated and therefore either remain stable or quickly fragment.    
\citet{rowther2024b} report that the shock heating is negligible compared to the stellar irradiation for a \solmass{0.1} disc around a \solmass{1} star at 3~Myr, meaning gravitoturbulence does not drive disc evolution for those parameters. As a result, studies have recently looked at whether a self-regulated state driven instead by infall is possible \citep{longarini2025b}. For the more optically thin discs with a low surface density and $L_* \gtrsim$ \sollum{1}, we agree that the stellar irradiation is indeed dominant. However, we do find that more massive discs around young protostars can develop a self-regulated state.

The longest we were able to follow the disc evolution for was 20 orbits ($\sim 7000$ years at 50~au) and several discs show sustained GI-driven spiral structure over this time (see Fig.~\ref{fig:Qminvt}). Furthermore, we see evidence of the transient spiral structure as predicted for a regime of marginal stability \citep{clarke2009}, for example in the simulation with $M_*=$~\solmass{1}, $R=50$~au, $\mu=0.3$ and see Fig.~\ref{fig:Qminvt}. However, the periodic variation in gravitational stress does not appear to cause distinctive variations in the accretion rate (Fig.~\ref{fig:accretionrates},  \citealt{lodatorice2005}).




\subsection{Implications for modelling}
We have shown that the new methods (combined and modified Lombardi, \citealt{young2024}) produce similar results which differ from the original Stamatellos method. Moving forward, we recommend the modified Lombardi method as a suitable radiative cooling approximation. The $\beta$ cooling approximation has been extensively employed for simulations of GI so now we examine how well imposed values of $\beta_{\rm cool}$ reflect the physical state of these discs.

The simplest simulations assume that $\beta_{\rm cool}$ is constant with radius but \citet{mercer2018} demonstrated that constant $\beta_{\rm cool}$ with radius is generally not realistic. They showed that $\beta_{\rm cool}$ generally decreases with radius, but in discs with spiral arms $\beta_{\rm cool} \sim$ constant in outer discs. {\edit Our results support this finding for discs with moderate surface density (for example, the blue line in Fig.~\ref{fig:Q_tcool}), though the slope is still quite steep ($\beta_{\rm cool}\propto 2$) and these discs tend to show faint, flocculent spirals. As expected, the $\beta$-cooling approximation with the exponential form is only representative of a subset of disc parameters and we do need to treat the thermodynamics to be self-consistent.}

{\edit In the last couple of years interest has grown in simulating self-gravitating discs with radiative transfer. Numerical radiative transfer methods effectively model the transport of heat within the disc but there is still much uncertainty surrounding the best approach to incorporating cooling to the external environment and stellar irradiation because these quantities require imposed boundary conditions. Very recently \citet{ni2025GIRT} and \citet{xu2025} studied the development of GI using radiation hydrodynamics. In the former model, energy loss to the environment derives from the local opacity and exposed cell surfaces, and in the latter an analytical cooling rate is applied at the disc surface to couple the disc thermodynamics to the environment. In radiation hydrodynamics simulations such as these (and also those implementing a polytropic cooling approximation), the cooling is naturally self-consistent with the development of GI. With such simulations, gravitoturbulent discs are generally not dominated by a single low-$m$ spiral, and fragments seem to form with lower masses than in simulations with fixed cooling rates.
A key limitation of those studies is that they do not include stellar heating. However, like us, \citet{xu2025} also found that grand-design spirals may be uncommon in gravitoturbulent discs which perhaps suggests that this is linked to more realistic treatment of thermodynamics rather than the inclusion of stellar heating.}

With the modified Lombardi radiative cooling approximation we can simulate discs using realistic physics with much more modest computational resources than would be required for full radiative transfer. A parameter study like the one presented here would be unfeasible with full radiative transfer, which illustrates a clear benefit to the new method. The reduced computational cost also allows simulations to be run for longer, which is important since disc structures can take many orbits to develop. Based on our results, we recommend evolving simulations for at least 10 dynamical timescales, by which time all the simulations presented here had reached their longer term state.

{\edit While the approximate radiative transfer methods compared here perform well for self-gravitating discs, there are scenarios for which full radiative transfer would be required. The approach used here is not suitable for modelling structures that cause shadowing (like all codes that employ FLD). Similarly, for complex geometries such as misaligned disc components only ray-tracing radiative transfer would accurately capture the thermodynamics. At present, the method cannot treat local variations in dust density since it relies upon a column-averaged density and opacity. However, we note that recent published simulations that employ ray-tracing radiative transfer also use constant dust-to-gas ratio \citep[e.g.][]{rowther2024b,xu2025}.}

\subsection{Predictions for observations}

The simulations conducted in this study indicate that discs need to be at least $0.3 M_*$ for the gravitational instability to develop. Therefore, we would expect the GI to be active only very early in the lifetime of the disc unless there has been a substantial late delivery of material to the disc \citep{hall2019}. The work also indicates that discs around lower mass stars may remain stable against fragmentation to $M_{\rm disc} \gtrsim M_*$. 
Star formation simulations predict that large (>80 au to several 100 au) discs can form \citep[e.g.][]{bate2018,wurster2019,He2025Chong-Chong} and that
$M_{\rm disc} \gtrsim M_*$ \citep{bate2011,bate2018} soon after the protostar forms. However, as pointed out by \citet{whitworth2006}, if we don't observe massive discs it doesn't mean they don't form; they might just fragment quickly and have short lifetimes. Observationally, the sample of very young discs in \citet{maureira2025} contains several discs with $M_{\rm disc} > 0.1 M_*$ and in which $Q<1$ for some radial range. This demonstrates the existence of young discs with significant masses and more observations of similar discs may be forthcoming given the growing understanding of the problems of underestimating disc masses from observations, especially at the early, embedded stages.

Long-lived, extended, high-contrast spiral structures appear to be a less common effect of GI and so we don't expect to find many in observations. Low contrast, higher-$m$ spiral structures are more common but these may be resolved out or hidden due to high optical depths \citep{maureira2025}. It is possible that dust trapping in spiral arms might enhance the observability of these structures \citep{dipierro2015,cadman2020} and this will be an interesting avenue for future research.

\section{Conclusion}
\label{sec:conclusion}

In this paper we have compared the evolution of massive discs around young protostars in simulations with four different implementations of a radiative cooling approximation. We find that because the earlier Stamatellos and Lombardi methods overestimate the optical depth, the parameter space for which discs develop spiral arms and/or fragment changes compared to \citet{cadman2020a} and \citet{haworth2020} for example.

Even for 50 au discs, we find that the outer regions are optically thin and fragmentation is possible. For more extended discs of $R_{\rm{out}}=$ 100 and 200 au, the optical depth is sufficiently low for stellar irradiation to stabilise discs against fragmentation up to masses of $\gtrsim 0.4 M_*$. The large scale spiral arms thought to be characteristic of gravitational instability form only in discs around stars $\lesssim$~\solmass{0.3}, due to the low luminosity, and in discs with a high surface density. {\edit Our simulations indicate that discs with active gravitational instability, with $Q_{\rm min} \lesssim 1.5$, may only have a very faint spiral structure that is very unlikely to be detectable.} This could explain the dearth of observed discs with these features. Nevertheless, for parameters where spiral structures do form, we find that long-lived spirals can persist in a self-regulating state in weakly irradiated discs.

Because massive discs around the youngest protostars may be stable, there is a large reservoir of planetary building material with the potential for forming massive planets quickly as required by observations \citep[e.g.][]{nixon2018,gratton2019,currie2022}. Since a self-regulated state may persist for some time, this allows the possibility of the formation of planetary cores by direct collapse of the solid component \citep{rowther2024,rice2025}.

Since large disc masses and low stellar luminosities are required for the GI to develop, we expect GI to play a key role in driving the evolution of the youngest discs, of less than 1 Myr old. Now that more observations are targeting Class 0 protostars, we should soon be in a strong position to determine the extent to which GI shapes young discs and the beginning of planet formation.

\section*{Acknowledgements}

We thank the anonymous reviewer for their comments that helped to improve the manuscript. AKY is grateful for a UKRI Stephen Hawking Fellowship, a University of Leeds University Academic Fellowship, and a Warwick Prize Fellowship. AKY and KR are grateful for support from the UK STFC via grant ST/V000594/1. FM and AKY are grateful for funding from the Royal Society.
This work used the DiRAC Data Intensive service (DIaL2) at the University of Leicester, managed by the University of Leicester Research Computing Service on behalf of the STFC DiRAC HPC Facility (www.dirac.ac.uk). The DiRAC service at Leicester was funded by BEIS, UKRI and STFC capital funding and STFC operations grants. DiRAC is part of the UKRI Digital Research Infrastructure. AKY thanks the DiRAC RSEs for advice on speeding up {\sc phantom} at runtime. This work also made use of the Avon HPC cluster operated by the Scientific Computing Research Technology Platform at the University of Warwick and the Aire HPC cluster at the University of Leeds. This work made use of {\sc splash} \citep{price2007}, {\sc numpy} \citep{harris2020} and {\sc matplotlib} \citep{hunter2007}.

\section*{Data Availability}
{\sc phantom} is a public code and the version used in this work can be cloned from \url{https://github.com/alisonkyoung1/phantom}. The full set-up files and files for running the simulations will be made available on publication, as will selected data files for the key results in this paper.



\bibliographystyle{mnras}
\bibliography{refs} 

@article{ni2025GIRT,
	adsurl = {https://ui.adsabs.harvard.edu/abs/2025ApJ...995...96N},
	archiveprefix = {arXiv},
	author = {{Ni}, Yang and {Deng}, Hongping and {Bai}, Xue-Ning},
	doi = {10.3847/1538-4357/ae16a4},
	eid = {96},
	eprint = {2510.19915},
	journal = {\apj},
	month = dec,
	number = {1},
	pages = {96},
	primaryclass = {astro-ph.EP},
	title = {{Radiation Hydrodynamics of Self-gravitating Protoplanetary Disks. I. Direct Formation of Gas Giants via Disk Fragmentation}},
	volume = {995},
	year = 2025}

@article{cossins2010opacity,
	adsurl = {https://ui.adsabs.harvard.edu/abs/2010MNRAS.401.2587C},
	archiveprefix = {arXiv},
	author = {{Cossins}, Peter and {Lodato}, Giuseppe and {Clarke}, Cathie},
	doi = {10.1111/j.1365-2966.2009.15835.x},
	eprint = {0910.0850},
	journal = {\mnras},
	month = feb,
	number = {4},
	pages = {2587-2598},
	primaryclass = {astro-ph.GA},
	title = {{The effects of opacity on gravitational stability in protoplanetary discs}},
	volume = {401},
	year = 2010}

@article{zamponi2021,
	adsurl = {https://ui.adsabs.harvard.edu/abs/2021MNRAS.508.2583Z},
	archiveprefix = {arXiv},
	author = {{Zamponi}, Joaquin and {Maureira}, Mar{\'\i}a Jos{\'e} and {Zhao}, Bo and {Liu}, Hauyu Baobab and {Ilee}, John D. and {Forgan}, Duncan and {Caselli}, Paola},
	doi = {10.1093/mnras/stab2657},
	eprint = {2109.06497},
	journal = {\mnras},
	month = dec,
	number = {2},
	pages = {2583-2599},
	primaryclass = {astro-ph.SR},
	title = {{The young protostellar disc in IRAS 16293-2422 B is hot and shows signatures of gravitational instability}},
	volume = {508},
	year = 2021}

@article{takakuwa2024,
	adsurl = {https://ui.adsabs.harvard.edu/abs/2024ApJ...964...24T},
	archiveprefix = {arXiv},
	author = {{Takakuwa}, Shigehisa and {Saigo}, Kazuya and {Kido}, Miyu and {Ohashi}, Nagayoshi and {Tobin}, John J. and {J{\o}rgensen}, Jes K. and {Aikawa}, Yuri and {Aso}, Yusuke and {Gavino}, Sacha and {Han}, Ilseung and {Koch}, Patrick M. and {Kwon}, Woojin and {Lee}, Chang Won and {Lee}, Jeong-Eun and {Li}, Zhi-Yun and {Lin}, Zhe-Yu Daniel and {Looney}, Leslie W. and {Mori}, Shoji and {Sai}, Jinshi, Jinshi (Insa Choi) and {Sharma}, Rajeeb and {Sheehan}, Patrick D. and {Tomida}, Kengo and {Williams}, Jonathan P. and {Yamato}, Yoshihide and {Yen}, Hsi-Wei},
	doi = {10.3847/1538-4357/ad1f57},
	eid = {24},
	eprint = {2401.08722},
	journal = {\apj},
	month = mar,
	number = {1},
	pages = {24},
	primaryclass = {astro-ph.EP},
	title = {{Early Planet Formation in Embedded Disks (eDisk). XIV. Flared Dust Distribution and Viscous Accretion Heating of the Disk around R CrA IRS 7B-a}},
	volume = {964},
	year = 2024}

@article{vanthoff2020,
	adsurl = {https://ui.adsabs.harvard.edu/abs/2020ApJ...901..166V},
	archiveprefix = {arXiv},
	author = {{van't Hoff}, Merel L.~R. and {Harsono}, Daniel and {Tobin}, John J. and {Bosman}, Arthur D. and {van Dishoeck}, Ewine F. and {J{\o}rgensen}, Jes K. and {Miotello}, Anna and {Murillo}, Nadia M. and {Walsh}, Catherine},
	doi = {10.3847/1538-4357/abb1a2},
	eid = {166},
	eprint = {2008.08106},
	journal = {\apj},
	month = oct,
	number = {2},
	pages = {166},
	primaryclass = {astro-ph.SR},
	title = {{Temperature Structures of Embedded Disks: Young Disks in Taurus Are Warm}},
	volume = {901},
	year = 2020}

@article{van-t-hoff2018ab,
	adsurl = {http://adsabs.harvard.edu/abs/2018A%26A...615A..83V},
	archiveprefix = {arXiv},
	author = {{van 't Hoff}, M.~L.~R. and {Tobin}, J.~J. and {Harsono}, D. and {van Dishoeck}, E.~F.},
	doi = {10.1051/0004-6361/201732313},
	eid = {A83},
	eprint = {1803.04515},
	journal = {\aap},
	month = jul,
	pages = {A83},
	primaryclass = {astro-ph.SR},
	title = {{Unveiling the physical conditions of the youngest disks. A warm embedded disk in L1527}},
	volume = 615,
	year = 2018}

@article{xu2025,
	adsurl = {https://ui.adsabs.harvard.edu/abs/2025ApJ...986...92X},
	archiveprefix = {arXiv},
	author = {{Xu}, Wenrui and {Jiang}, Yan-Fei and {Kunz}, Matthew W. and {Stone}, James M.},
	doi = {10.3847/1538-4357/add14b},
	eid = {92},
	eprint = {2504.18751},
	journal = {\apj},
	month = jun,
	number = {1},
	pages = {92},
	primaryclass = {astro-ph.EP},
	title = {{Global Simulations of Gravitational Instability in Protostellar Disks with Full Radiation Transport. II. Locality of Gravitoturbulence, Clumpy Spirals, and Implications for Observable Substructure}},
	volume = {986},
	year = 2025}

@article{linkratter2016,
	adsurl = {https://ui.adsabs.harvard.edu/abs/2016ApJ...824...91L},
	archiveprefix = {arXiv},
	author = {{Lin}, Min-Kai and {Kratter}, Kaitlin M.},
	doi = {10.3847/0004-637X/824/2/91},
	eid = {91},
	eprint = {1603.01613},
	journal = {\apj},
	month = jun,
	number = {2},
	pages = {91},
	primaryclass = {astro-ph.EP},
	title = {{On the Gravitational Stability of Gravito-turbulent Accretion Disks}},
	volume = {824},
	year = 2016}

@article{mamatsashvili2010,
	adsurl = {https://ui.adsabs.harvard.edu/abs/2010MNRAS.406.2050M},
	archiveprefix = {arXiv},
	author = {{Mamatsashvili}, G.~R. and {Rice}, W.~K.~M.},
	doi = {10.1111/j.1365-2966.2010.16825.x},
	eprint = {1004.1662},
	journal = {\mnras},
	month = aug,
	number = {3},
	pages = {2050-2064},
	primaryclass = {astro-ph.EP},
	title = {{Axisymmetric modes in vertically stratified self-gravitating discs}},
	volume = {406},
	year = 2010}

@article{deng2021,
	adsurl = {https://ui.adsabs.harvard.edu/abs/2021NatAs...5..440D},
	archiveprefix = {arXiv},
	author = {{Deng}, Hongping and {Mayer}, Lucio and {Helled}, Ravit},
	doi = {10.1038/s41550-020-01297-6},
	eprint = {2101.01331},
	journal = {Nature Astronomy},
	month = feb,
	pages = {440-444},
	primaryclass = {astro-ph.EP},
	title = {{Formation of intermediate-mass planets via magnetically controlled disk fragmentation}},
	volume = {5},
	year = 2021}

@article{meru2015,
	adsurl = {https://ui.adsabs.harvard.edu/abs/2015MNRAS.454.2529M},
	archiveprefix = {arXiv},
	author = {{Meru}, Farzana},
	doi = {10.1093/mnras/stv2128},
	eprint = {1509.03635},
	journal = {\mnras},
	month = dec,
	number = {3},
	pages = {2529-2538},
	primaryclass = {astro-ph.EP},
	title = {{Triggered fragmentation in self-gravitating discs: forming fragments at small radii}},
	volume = {454},
	year = 2015}

@article{matzner2005,
	adsurl = {https://ui.adsabs.harvard.edu/abs/2005ApJ...628..817M},
	archiveprefix = {arXiv},
	author = {{Matzner}, Christopher D. and {Levin}, Yuri},
	doi = {10.1086/430813},
	eprint = {astro-ph/0408525},
	journal = {\apj},
	month = aug,
	number = {2},
	pages = {817-831},
	primaryclass = {astro-ph},
	title = {{Protostellar Disks: Formation, Fragmentation, and the Brown Dwarf Desert}},
	volume = {628},
	year = 2005}

@article{meru2010,
	adsurl = {https://ui.adsabs.harvard.edu/abs/2010MNRAS.406.2279M},
	archiveprefix = {arXiv},
	author = {{Meru}, Farzana and {Bate}, Matthew R.},
	doi = {10.1111/j.1365-2966.2010.16867.x},
	eprint = {1004.3766},
	journal = {\mnras},
	month = aug,
	number = {4},
	pages = {2279-2288},
	primaryclass = {astro-ph.EP},
	title = {{Exploring the conditions required to form giant planets via gravitational instability in massive protoplanetary discs}},
	volume = {406},
	year = 2010}

@article{bertin1998,
	adsurl = {https://ui.adsabs.harvard.edu/abs/1988A&A...195..105B},
	author = {{Bertin}, G. and {Romeo}, A.~B.},
	journal = {\aap},
	month = apr,
	pages = {105-113},
	title = {{Global spiral modes in stellar disks containing gas.}},
	volume = {195},
	year = 1988}

@article{vandervoort1970,
	adsurl = {https://ui.adsabs.harvard.edu/abs/1970ApJ...161...87V},
	author = {{Vandervoort}, Peter O.},
	doi = {10.1086/150514},
	journal = {\apj},
	month = jul,
	pages = {87},
	title = {{Density Waves in a Highly Flattened, Rapidly Rotating Galaxy}},
	volume = {161},
	year = 1970}

@article{safronov1960,
	adsurl = {https://ui.adsabs.harvard.edu/abs/1960AnAp...23..979S},
	author = {{Safronov}, V.~S.},
	journal = {Annales d'Astrophysique},
	month = feb,
	pages = {979},
	title = {{On the gravitational instability in flattened systems with axial symmetry and non-uniform rotation}},
	volume = {23},
	year = 1960}

@article{goldreich1965,
	adsurl = {https://ui.adsabs.harvard.edu/abs/1965MNRAS.130...97G},
	author = {{Goldreich}, P. and {Lynden-Bell}, D.},
	doi = {10.1093/mnras/130.2.97},
	journal = {\mnras},
	month = jan,
	pages = {97},
	title = {{I. Gravitational stability of uniformly rotating disks}},
	volume = {130},
	year = 1965}

@article{dipierro2015,
	adsurl = {https://ui.adsabs.harvard.edu/abs/2015MNRAS.451..974D},
	archiveprefix = {arXiv},
	author = {{Dipierro}, Giovanni and {Pinilla}, Paola and {Lodato}, Giuseppe and {Testi}, Leonardo},
	doi = {10.1093/mnras/stv970},
	eprint = {1504.08099},
	journal = {\mnras},
	month = jul,
	number = {1},
	pages = {974-986},
	primaryclass = {astro-ph.SR},
	title = {{Dust trapping by spiral arms in gravitationally unstable protostellar discs}},
	volume = {451},
	year = 2015}

@article{cadman2020,
	adsurl = {https://ui.adsabs.harvard.edu/abs/2020MNRAS.498.4256C},
	archiveprefix = {arXiv},
	author = {{Cadman}, James and {Hall}, Cassandra and {Rice}, Ken and {Harries}, Tim J. and {Klaassen}, Pamela D.},
	doi = {10.1093/mnras/staa2596},
	eprint = {2008.09826},
	journal = {\mnras},
	month = nov,
	number = {3},
	pages = {4256-4271},
	primaryclass = {astro-ph.EP},
	title = {{The observational impact of dust trapping in self-gravitating discs}},
	volume = {498},
	year = 2020}

@article{lynden-bell1972,
	adsurl = {https://ui.adsabs.harvard.edu/abs/1972MNRAS.157....1L},
	author = {{Lynden-Bell}, D. and {Kalnajs}, A.~J.},
	doi = {10.1093/mnras/157.1.1},
	journal = {\mnras},
	month = jan,
	pages = {1},
	title = {{On the generating mechanism of spiral structure}},
	volume = {157},
	year = 1972}

@article{lee2025,
	adsurl = {https://ui.adsabs.harvard.edu/abs/2025MNRAS.544L..18L},
	archiveprefix = {arXiv},
	author = {{Lee}, Hans and {Nayakshin}, Sergei and {Booth}, Richard A.},
	doi = {10.1093/mnrasl/slaf096},
	eprint = {2509.09305},
	journal = {\mnras},
	month = nov,
	number = {1},
	pages = {L18-L23},
	primaryclass = {astro-ph.EP},
	title = {{Dust growth and planet formation by disc fragmentation}},
	volume = {544},
	year = 2025}

@article{schib2025,
	adsurl = {https://ui.adsabs.harvard.edu/abs/2025A&A...704A..28S},
	archiveprefix = {arXiv},
	author = {{Schib}, O. and {Mordasini}, C. and {Emsenhuber}, A. and {Helled}, R.},
	doi = {10.1051/0004-6361/202556261},
	eid = {A28},
	eprint = {2510.02437},
	journal = {\aap},
	month = dec,
	pages = {A28},
	primaryclass = {astro-ph.EP},
	title = {{DIPSY: A new Disc Instability Population SYnthesis: II. The Populations of Companions Formed Through Disc Instability}},
	volume = {704},
	year = 2025}

@ARTICLE{hall2019,
       author = {{Hall}, Cassandra and {Dong}, Ruobing and {Rice}, Ken and {Harries}, Tim J. and {Najita}, Joan and {Alexander}, Richard and {Brittain}, Sean},
        title = "{The Temporal Requirements of Directly Observing Self-gravitating Spiral Waves in Protoplanetary Disks with ALMA}",
      journal = {\apj},
         year = 2019,
        month = feb,
       volume = {871},
       number = {2},
          eid = {228},
        pages = {228},
          doi = {10.3847/1538-4357/aafac2},
archivePrefix = {arXiv},
       eprint = {1901.02407},
 primaryClass = {astro-ph.EP},
       adsurl = {https://ui.adsabs.harvard.edu/abs/2019ApJ...871..228H}
}

@article{rafikov2005,
	adsurl = {https://ui.adsabs.harvard.edu/abs/2005ApJ...621L..69R},
	archiveprefix = {arXiv},
	author = {{Rafikov}, Roman R.},
	doi = {10.1086/428899},
	eprint = {astro-ph/0406469},
	journal = {\apjl},
	month = mar,
	number = {1},
	pages = {L69-L72},
	primaryclass = {astro-ph},
	title = {{Can Giant Planets Form by Direct Gravitational Instability?}},
	volume = {621},
	year = 2005}

@article{mayer2004,
	adsurl = {https://ui.adsabs.harvard.edu/abs/2004ApJ...609.1045M},
	archiveprefix = {arXiv},
	author = {{Mayer}, Lucio and {Quinn}, Thomas and {Wadsley}, James and {Stadel}, Joachim},
	doi = {10.1086/421288},
	eprint = {astro-ph/0310771},
	journal = {\apj},
	month = jul,
	number = {2},
	pages = {1045-1064},
	primaryclass = {astro-ph},
	title = {{The Evolution of Gravitationally Unstable Protoplanetary Disks: Fragmentation and Possible Giant Planet Formation}},
	volume = {609},
	year = 2004}

@article{boss1998,
	adsurl = {https://ui.adsabs.harvard.edu/abs/1998ApJ...503..923B},
	author = {{Boss}, Alan P.},
	doi = {10.1086/306036},
	journal = {\apj},
	month = aug,
	number = {2},
	pages = {923-937},
	title = {{Evolution of the Solar Nebula. IV. Giant Gaseous Protoplanet Formation}},
	volume = {503},
	year = 1998}

@article{rowther2024,
	adsurl = {https://ui.adsabs.harvard.edu/abs/2024MNRAS.528.2490R},
	archiveprefix = {arXiv},
	author = {{Rowther}, Sahl and {Nealon}, Rebecca and {Meru}, Farzana and {Wurster}, James and {Aly}, Hossam and {Alexander}, Richard and {Rice}, Ken and {Booth}, Richard A.},
	doi = {10.1093/mnras/stae167},
	eprint = {2401.09380},
	journal = {\mnras},
	month = feb,
	number = {2},
	pages = {2490-2500},
	title = {{The role of drag and gravity on dust concentration in a gravitationally unstable disc}},
	volume = {528},
	year = 2024}

@article{shakura1973,
	author = {{Shakura}, N.~,I.~ and {Sunyaev}, R.~A.~},
	journal = {A\&A},
	month = {6},
	pages = {337},
	title = {Black holes in binary systems. Observational appearance.},
	volume = {24},
	year = {1973}}

@ARTICLE{rowther2023hiding,
       author = {{Rowther}, Sahl and {Nealon}, Rebecca and {Meru}, Farzana},
        title = "{Continuing to hide signatures of gravitational instability in protoplanetary discs with planets}",
      journal = {\mnras},
         year = 2023,
        month = jan,
       volume = {518},
       number = {1},
        pages = {763-773},
          doi = {10.1093/mnras/stac3106},
archivePrefix = {arXiv},
       eprint = {2210.17454},
 primaryClass = {astro-ph.EP},
       adsurl = {https://ui.adsabs.harvard.edu/abs/2023MNRAS.518..763R}
}

@article{lodatorice2004,
	adsurl = {https://ui.adsabs.harvard.edu/abs/2004MNRAS.351..630L},
	archiveprefix = {arXiv},
	author = {{Lodato}, G. and {Rice}, W.~K.~M.},
	doi = {10.1111/j.1365-2966.2004.07811.x},
	eprint = {astro-ph/0403185},
	journal = {\mnras},
	month = jun,
	number = {2},
	pages = {630-642},
	primaryclass = {astro-ph},
	title = {{Testing the locality of transport in self-gravitating accretion discs}},
	volume = {351},
	year = 2004}

@article{gratton2019,
	adsurl = {https://ui.adsabs.harvard.edu/abs/2019A&A...623A.140G},
	archiveprefix = {arXiv},
	author = {{Gratton}, R. and {Ligi}, R. and {Sissa}, E. and {Desidera}, S. and {Mesa}, D. and {Bonnefoy}, M. and {Chauvin}, G. and {Cheetham}, A. and {Feldt}, M. and {Lagrange}, A.~M. and {Langlois}, M. and {Meyer}, M. and {Vigan}, A. and {Boccaletti}, A. and {Janson}, M. and {Lazzoni}, C. and {Zurlo}, A. and {De Boer}, J. and {Henning}, T. and {D'Orazi}, V. and {Gluck}, L. and {Madec}, F. and {Jaquet}, M. and {Baudoz}, P. and {Fantinel}, D. and {Pavlov}, A. and {Wildi}, F.},
	doi = {10.1051/0004-6361/201834760},
	eid = {A140},
	eprint = {1901.06555},
	journal = {\aap},
	month = mar,
	pages = {A140},
	primaryclass = {astro-ph.EP},
	title = {{Blobs, spiral arms, and a possible planet around HD 169142}},
	volume = {623},
	year = 2019}

@article{currie2022,
	adsurl = {https://ui.adsabs.harvard.edu/abs/2022NatAs...6..751C},
	archiveprefix = {arXiv},
	author = {{Currie}, Thayne and {Lawson}, Kellen and {Schneider}, Glenn and {Lyra}, Wladimir and {Wisniewski}, John and {Grady}, Carol and {Guyon}, Olivier and {Tamura}, Motohide and {Kotani}, Takayuki and {Kawahara}, Hajime and {Brandt}, Timothy and {Uyama}, Taichi and {Muto}, Takayuki and {Dong}, Ruobing and {Kudo}, Tomoyuki and {Hashimoto}, Jun and {Fukagawa}, Misato and {Wagner}, Kevin and {Lozi}, Julien and {Chilcote}, Jeffrey and {Tobin}, Taylor and {Groff}, Tyler and {Ward-Duong}, Kimberly and {Januszewski}, William and {Norris}, Barnaby and {Tuthill}, Peter and {van der Marel}, Nienke and {Sitko}, Michael and {Deo}, Vincent and {Vievard}, Sebastien and {Jovanovic}, Nemanja and {Martinache}, Frantz and {Skaf}, Nour},
	doi = {10.1038/s41550-022-01634-x},
	eprint = {2204.00633},
	journal = {Nature Astronomy},
	month = apr,
	pages = {751-759},
	primaryclass = {astro-ph.EP},
	title = {{Images of embedded Jovian planet formation at a wide separation around AB Aurigae}},
	volume = {6},
	year = 2022}

@article{laughlin1996,
	adsurl = {https://ui.adsabs.harvard.edu/abs/1996ApJ...456..279L},
	author = {{Laughlin}, Gregory and {Rozyczka}, Michal},
	doi = {10.1086/176648},
	journal = {\apj},
	month = jan,
	pages = {279},
	title = {{The Effect of Gravitational Instabilities on Protostellar Disks}},
	volume = {456},
	year = 1996}

@article{chiang1997,
	adsurl = {https://ui.adsabs.harvard.edu/abs/1997ApJ...490..368C},
	archiveprefix = {arXiv},
	author = {{Chiang}, E.~I. and {Goldreich}, P.},
	doi = {10.1086/304869},
	eprint = {astro-ph/9706042},
	journal = {\apj},
	month = nov,
	number = {1},
	pages = {368-376},
	primaryclass = {astro-ph},
	title = {{Spectral Energy Distributions of T Tauri Stars with Passive Circumstellar Disks}},
	volume = {490},
	year = 1997}

@article{hunter2007,
	author = {{Hunter, J. D.}},
	journal = {{Computing in Science \& Engineering}},
	number = {3},
	pages = {90--95},
	title = {{Matplotlib: A 2D graphics environment}},
	volume = {9},
	year = {2007}}

@article{harris2020,
	author = {Charles R. Harris and K. Jarrod Millman and St{'{e}}fan J. van der Walt and Ralf Gommers and Pauli Virtanen and David Cournapeau and Eric Wieser and Julian Taylor and Sebastian Berg and Nathaniel J. Smith and Robert Kern and Matti Picus and Stephan Hoyer and Marten H. van Kerkwijk and Matthew Brett and Allan Haldane and Jaime Fern{'{a}}ndez del R{'{\i}}o and Mark Wiebe and Pearu Peterson and Pierre G{'{e}}rard-Marchant and Kevin Sheppard and Tyler Reddy and Warren Weckesser and Hameer Abbasi and Christoph Gohlke and Travis E. Oliphant},
	journal = {Nature},
	month = {sep},
	number = {7825},
	pages = {357--362},
	title = {Array programming with {NumPy}},
	volume = {585},
	year = {2020}}

@article{price2007,
	adsurl = {http://adsabs.harvard.edu/abs/2007PASA...24..159P},
	archiveprefix = {arXiv},
	author = {{Price}, D.~J.},
	doi = {10.1071/AS07022},
	eprint = {0709.0832},
	journal = {\pasa},
	month = oct,
	pages = {159-173},
	title = {{splash: An Interactive Visualisation Tool for Smoothed Particle Hydrodynamics Simulations}},
	volume = 24,
	year = 2007}

@article{wurster2019,
	adsurl = {https://ui.adsabs.harvard.edu/abs/2019MNRAS.489.1719W},
	archiveprefix = {arXiv},
	author = {{Wurster}, James and {Bate}, Matthew R. and {Price}, Daniel J.},
	doi = {10.1093/mnras/stz2215},
	eprint = {1908.03241},
	journal = {\mnras},
	month = oct,
	number = {2},
	pages = {1719-1741},
	primaryclass = {astro-ph.SR},
	title = {{There is no magnetic braking catastrophe: low-mass star cluster and protostellar disc formation with non-ideal magnetohydrodynamics}},
	volume = {489},
	year = 2019}

@article{bate2018,
	adsurl = {https://ui.adsabs.harvard.edu/abs/2018MNRAS.475.5618B},
	archiveprefix = {arXiv},
	author = {{Bate}, Matthew R.},
	doi = {10.1093/mnras/sty169},
	eprint = {1801.07721},
	journal = {\mnras},
	month = apr,
	number = {4},
	pages = {5618-5658},
	primaryclass = {astro-ph.SR},
	title = {{On the diversity and statistical properties of protostellar discs}},
	volume = {475},
	year = 2018}

@article{bate2011,
	adsurl = {https://ui.adsabs.harvard.edu/abs/2011MNRAS.417.2036B},
	archiveprefix = {arXiv},
	author = {{Bate}, Matthew R.},
	doi = {10.1111/j.1365-2966.2011.19386.x},
	eprint = {1108.0009},
	journal = {\mnras},
	month = nov,
	number = {3},
	pages = {2036-2056},
	primaryclass = {astro-ph.SR},
	title = {{Collapse of a molecular cloud core to stellar densities: the formation and evolution of pre-stellar discs}},
	volume = {417},
	year = 2011}

@article{cai2008,
	adsurl = {https://ui.adsabs.harvard.edu/abs/2008ApJ...673.1138C},
	archiveprefix = {arXiv},
	author = {{Cai}, Kai and {Durisen}, Richard H. and {Boley}, Aaron C. and {Pickett}, Megan K. and {Mej{\'\i}a}, Annie C.},
	doi = {10.1086/524101},
	eprint = {0706.4046},
	journal = {\apj},
	month = feb,
	number = {2},
	pages = {1138-1153},
	primaryclass = {astro-ph},
	title = {{The Thermal Regulation of Gravitational Instabilities in Protoplanetary Disks. IV. Simulations with Envelope Irradiation}},
	volume = {673},
	year = 2008}

@article{forgan2018,
	adsurl = {https://ui.adsabs.harvard.edu/abs/2018MNRAS.474.5036F},
	archiveprefix = {arXiv},
	author = {{Forgan}, D.~H. and {Hall}, C. and {Meru}, F. and {Rice}, W.~K.~M.},
	doi = {10.1093/mnras/stx2870},
	eprint = {1711.01133},
	journal = {\mnras},
	month = mar,
	number = {4},
	pages = {5036-5048},
	primaryclass = {astro-ph.EP},
	title = {{Towards a population synthesis model of self-gravitating disc fragmentation and tidal downsizing II: the effect of fragment-fragment interactions}},
	volume = {474},
	year = 2018}

@article{forgan2013,
	adsurl = {https://ui.adsabs.harvard.edu/abs/2013MNRAS.430.2082F},
	archiveprefix = {arXiv},
	author = {{Forgan}, Duncan and {Rice}, Ken},
	doi = {10.1093/mnras/stt032},
	eprint = {1301.1151},
	journal = {\mnras},
	month = apr,
	number = {3},
	pages = {2082-2089},
	primaryclass = {astro-ph.SR},
	title = {{The effect of irradiation on the Jeans mass in fragmenting self-gravitating protostellar discs}},
	volume = {430},
	year = 2013}

@article{whitworth2006,
	adsurl = {https://ui.adsabs.harvard.edu/abs/2006A&A...458..817W},
	archiveprefix = {arXiv},
	author = {{Whitworth}, A.~P. and {Stamatellos}, D.},
	doi = {10.1051/0004-6361:20065806},
	eprint = {astro-ph/0610039},
	journal = {\aap},
	month = nov,
	number = {3},
	pages = {817-829},
	primaryclass = {astro-ph},
	title = {{The minimum mass for star formation, and the origin of binary brown dwarfs}},
	volume = {458},
	year = 2006}

@inproceedings{stamatellos2009,
	adsurl = {https://ui.adsabs.harvard.edu/abs/2009AIPC.1094..557S},
	archiveprefix = {arXiv},
	author = {{Stamatellos}, Dimitris and {Whitworth}, Anthony P.},
	booktitle = {15th Cambridge Workshop on Cool Stars, Stellar Systems, and the Sun},
	doi = {10.1063/1.3099172},
	editor = {{Stempels}, Eric},
	eprint = {0809.5042},
	month = feb,
	pages = {557-560},
	primaryclass = {astro-ph},
	publisher = {AIP},
	series = {American Institute of Physics Conference Series},
	title = {{The formation of brown dwarfs and low-mass stars by disc fragmentation}},
	volume = {1094},
	year = 2009}

@article{stamatellos2009thermo,
	adsurl = {https://ui.adsabs.harvard.edu/abs/2009MNRAS.400.1563S},
	archiveprefix = {arXiv},
	author = {{Stamatellos}, Dimitris and {Whitworth}, Anthony P.},
	doi = {10.1111/j.1365-2966.2009.15564.x},
	eprint = {0908.2247},
	journal = {\mnras},
	month = dec,
	number = {3},
	pages = {1563-1573},
	primaryclass = {astro-ph.GA},
	title = {{The role of thermodynamics in disc fragmentation}},
	volume = {400},
	year = 2009}

@article{lodatorice2005,
	adsurl = {https://ui.adsabs.harvard.edu/abs/2005MNRAS.358.1489L},
	archiveprefix = {arXiv},
	author = {{Lodato}, G. and {Rice}, W.~K.~M.},
	doi = {10.1111/j.1365-2966.2005.08875.x},
	eprint = {astro-ph/0501638},
	journal = {\mnras},
	month = apr,
	number = {4},
	pages = {1489-1500},
	primaryclass = {astro-ph},
	title = {{Testing the locality of transport in self-gravitating accretion discs - II. The massive disc case}},
	volume = {358},
	year = 2005}

@article{rice2015,
	adsurl = {https://ui.adsabs.harvard.edu/abs/2015MNRAS.454.1940R},
	archiveprefix = {arXiv},
	author = {{Rice}, Ken and {Lopez}, Eric and {Forgan}, Duncan and {Biller}, Beth},
	doi = {10.1093/mnras/stv1997},
	eprint = {1508.06528},
	journal = {\mnras},
	month = dec,
	number = {2},
	pages = {1940-1947},
	primaryclass = {astro-ph.EP},
	title = {{Disc fragmentation rarely forms planetary-mass objects}},
	volume = {454},
	year = 2015}

@article{rice2016,
	adsurl = {https://ui.adsabs.harvard.edu/abs/2016PASA...33...12R},
	archiveprefix = {arXiv},
	author = {{Rice}, Ken},
	doi = {10.1017/pasa.2016.12},
	eid = {e012},
	eprint = {1602.08390},
	journal = {\pasa},
	month = mar,
	pages = {e012},
	primaryclass = {astro-ph.SR},
	title = {{The Evolution of Self-Gravitating Accretion Discs}},
	volume = {33},
	year = 2016}

@article{maureira2025,
	adsurl = {https://ui.adsabs.harvard.edu/abs/2026A&A...705A..96M},
	archiveprefix = {arXiv},
	author = {{Maureira}, M.~J. and {Pineda}, J.~E. and {Liu}, H.~B. and {Caselli}, P. and {Chandler}, C. and {Testi}, L. and {Johnstone}, D. and {Segura-Cox}, D. and {Loinard}, L. and {Bianchi}, E. and {Codella}, C. and {Miotello}, A. and {Podio}, L. and {Cacciapuoti}, L. and {Oya}, Y. and {Lopez-Sepulcre}, A. and {Sakai}, N. and {Zhang}, Z. and {Cuello}, N. and {Ohashi}, S. and {Aikawa}, Y. and {Sabatini}, G. and {Zhang}, Y. and {Ceccarelli}, C. and {Yamamoto}, S.},
	doi = {10.1051/0004-6361/202556063},
	eid = {A96},
	eprint = {2510.19635},
	journal = {\aap},
	month = jan,
	pages = {A96},
	primaryclass = {astro-ph.SR},
	title = {{FAUST: XXVIII. High-resolution ALMA observations of Class 0/I disks: Structure, optical depths, and temperatures}},
	volume = {705},
	year = 2026}

@article{laughlin1994,
	adsurl = {https://ui.adsabs.harvard.edu/abs/1994ApJ...436..335L},
	author = {{Laughlin}, Gregory and {Bodenheimer}, Peter},
	doi = {10.1086/174909},
	journal = {\apj},
	month = nov,
	pages = {335},
	title = {{Nonaxisymmetric Evolution in Protostellar Disks}},
	volume = {436},
	year = 1994}

@article{jorgensen2009prosac,
	adsurl = {https://ui.adsabs.harvard.edu/abs/2009A&A...507..861J},
	archiveprefix = {arXiv},
	author = {{J{\o}rgensen}, J.~K. and {van Dishoeck}, E.~F. and {Visser}, R. and {Bourke}, T.~L. and {Wilner}, D.~J. and {Lommen}, D. and {Hogerheijde}, M.~R. and {Myers}, P.~C.},
	doi = {10.1051/0004-6361/200912325},
	eprint = {0909.3386},
	journal = {\aap},
	month = nov,
	number = {2},
	pages = {861-879},
	primaryclass = {astro-ph.SR},
	title = {{PROSAC: a submillimeter array survey of low-mass protostars. II. The mass evolution of envelopes, disks, and stars from the Class 0 through I stages}},
	volume = {507},
	year = 2009}

@article{williams2011araa,
	adsurl = {https://ui.adsabs.harvard.edu/abs/2011ARA&A..49...67W},
	archiveprefix = {arXiv},
	author = {{Williams}, Jonathan P. and {Cieza}, Lucas A.},
	doi = {10.1146/annurev-astro-081710-102548},
	eprint = {1103.0556},
	journal = {\araa},
	month = sep,
	number = {1},
	pages = {67-117},
	primaryclass = {astro-ph.GA},
	title = {{Protoplanetary Disks and Their Evolution}},
	volume = {49},
	year = 2011}

@article{longarini2025b,
	adsurl = {https://ui.adsabs.harvard.edu/abs/2025MNRAS.541.1145L},
	archiveprefix = {arXiv},
	author = {{Longarini}, Cristiano and {Price}, Daniel J. and {Kratter}, Kaitlin M. and {Lodato}, Giuseppe and {Clarke}, Cathie J.},
	doi = {10.1093/mnras/staf1018},
	eprint = {2506.13701},
	journal = {\mnras},
	month = aug,
	number = {2},
	pages = {1145-1163},
	primaryclass = {astro-ph.EP},
	title = {{Infall-driven gravitational instability in accretion discs}},
	volume = {541},
	year = 2025}

@article{leedham2025,
	adsurl = {https://ui.adsabs.harvard.edu/abs/2025MNRAS.539.2780L},
	archiveprefix = {arXiv},
	author = {{Leedham}, Caitriona S. and {Booth}, Richard A. and {Clarke}, Cathie J.},
	doi = {10.1093/mnras/staf644},
	eprint = {2503.22667},
	journal = {\mnras},
	month = may,
	number = {3},
	pages = {2780-2789},
	primaryclass = {astro-ph.EP},
	title = {{Effect of irradiation model on 2D hydrodynamic simulations of self-gravitating protoplanetary discs}},
	volume = {539},
	year = 2025}

@article{rowther2024b,
	adsurl = {https://ui.adsabs.harvard.edu/abs/2024MNRAS.534.2277R},
	archiveprefix = {arXiv},
	author = {{Rowther}, Sahl and {Price}, Daniel J. and {Pinte}, Christophe and {Nealon}, Rebecca and {Meru}, Farzana and {Alexander}, Richard},
	doi = {10.1093/mnras/stae2167},
	eprint = {2409.10765},
	journal = {\mnras},
	month = nov,
	number = {3},
	pages = {2277-2285},
	primaryclass = {astro-ph.EP},
	title = {{Short-lived gravitational instability in isolated irradiated discs}},
	volume = {534},
	year = 2024}

@article{rice2011,
	adsurl = {https://ui.adsabs.harvard.edu/abs/2011MNRAS.418.1356R},
	archiveprefix = {arXiv},
	author = {{Rice}, W.~K.~M. and {Armitage}, P.~J. and {Mamatsashvili}, G.~R. and {Lodato}, G. and {Clarke}, C.~J.},
	doi = {10.1111/j.1365-2966.2011.19586.x},
	eprint = {1108.1194},
	journal = {\mnras},
	month = dec,
	number = {2},
	pages = {1356-1362},
	primaryclass = {astro-ph.SR},
	title = {{Stability of self-gravitating discs under irradiation}},
	volume = {418},
	year = 2011}

@article{rice2025,
	adsurl = {https://ui.adsabs.harvard.edu/abs/2025MNRAS.539.3421R},
	archiveprefix = {arXiv},
	author = {{Rice}, Ken and {Baehr}, Hans and {Young}, Alison K. and {Booth}, Richard and {Rowther}, Sahl and {Meru}, Farzana and {Hall}, Cassandra and {Koval}, Adam},
	doi = {10.1093/mnras/staf714},
	eprint = {2505.00363},
	journal = {\mnras},
	month = jun,
	number = {4},
	pages = {3421-3435},
	primaryclass = {astro-ph.EP},
	title = {{Dust density enhancements and the direct formation of planetary cores in gravitationally unstable discs}},
	volume = {539},
	year = 2025}

@article{He2025Chong-Chong,
	adsurl = {https://ui.adsabs.harvard.edu/abs/2025MNRAS.540..175H},
	archiveprefix = {arXiv},
	author = {{He}, Chong-Chong and {Ricotti}, Massimo},
	doi = {10.1093/mnras/staf743},
	eprint = {2403.09779},
	journal = {\mnras},
	month = jun,
	number = {1},
	pages = {175-189},
	primaryclass = {astro-ph.GA},
	title = {{Formation of large circumstellar discs in multiscale, ideal-MHD simulations of magnetically critical, massive pre-stellar cores}},
	volume = {540},
	year = 2025}

@article{rowther2020,
	adsurl = {https://ui.adsabs.harvard.edu/abs/2020MNRAS.496.1598R},
	archiveprefix = {arXiv},
	author = {{Rowther}, Sahl and {Meru}, Farzana},
	doi = {10.1093/mnras/staa1590},
	eprint = {2006.03077},
	journal = {\mnras},
	month = aug,
	number = {2},
	pages = {1598-1609},
	primaryclass = {astro-ph.EP},
	title = {{Planet migration in self-gravitating discs: survival of planets}},
	volume = {496},
	year = 2020}

@article{forgan2009,
	adsurl = {https://ui.adsabs.harvard.edu/abs/2009MNRAS.394..882F},
	archiveprefix = {arXiv},
	author = {{Forgan}, Duncan and {Rice}, Ken and {Stamatellos}, Dimitris and {Whitworth}, Anthony},
	doi = {10.1111/j.1365-2966.2008.14373.x},
	eprint = {0812.0304},
	journal = {\mnras},
	month = apr,
	number = {2},
	pages = {882-891},
	primaryclass = {astro-ph},
	title = {{Introducing a hybrid radiative transfer method for smoothed particle hydrodynamics}},
	volume = {394},
	year = 2009}

@article{mercer2020,
	adsurl = {https://ui.adsabs.harvard.edu/abs/2020A&A...633A.116M},
	archiveprefix = {arXiv},
	author = {{Mercer}, Anthony and {Stamatellos}, Dimitris},
	doi = {10.1051/0004-6361/201936954},
	eid = {A116},
	eprint = {2001.10062},
	journal = {\aap},
	month = jan,
	pages = {A116},
	primaryclass = {astro-ph.SR},
	title = {{Planet formation around M dwarfs via disc instability. Fragmentation conditions and protoplanet properties}},
	volume = {633},
	year = 2020}

@article{dotter2016,
	adsurl = {https://ui.adsabs.harvard.edu/abs/2016ApJS..222....8D},
	archiveprefix = {arXiv},
	author = {{Dotter}, Aaron},
	doi = {10.3847/0067-0049/222/1/8},
	eid = {8},
	eprint = {1601.05144},
	journal = {\apjs},
	month = jan,
	number = {1},
	pages = {8},
	primaryclass = {astro-ph.SR},
	title = {{MESA Isochrones and Stellar Tracks (MIST) 0: Methods for the Construction of Stellar Isochrones}},
	volume = {222},
	year = 2016}

@article{bate1995aa,
	adsurl = {https://ui.adsabs.harvard.edu/#abs/1995MNRAS.277..362B},
	author = {{Bate}, Matthew R. and {Bonnell}, Ian A. and {Price}, Nigel M.},
	doi = {10.1093/mnras/277.2.362},
	journal = {\mnras},
	month = Nov,
	pages = {362-376},
	title = {{Modelling accretion in protobinary systems}},
	volume = {277},
	year = 1995}

@article{cullen2010,
	adsurl = {https://ui.adsabs.harvard.edu/abs/2010MNRAS.408..669C},
	archiveprefix = {arXiv},
	author = {{Cullen}, Lee and {Dehnen}, Walter},
	doi = {10.1111/j.1365-2966.2010.17158.x},
	eprint = {1006.1524},
	journal = {\mnras},
	month = oct,
	number = {2},
	pages = {669-683},
	primaryclass = {astro-ph.IM},
	title = {{Inviscid smoothed particle hydrodynamics}},
	volume = {408},
	year = 2010}

@article{price2018aa,
	adsurl = {https://ui.adsabs.harvard.edu/#abs/2018PASA...35...31P},
	author = {{Price}, Daniel J. and {Wurster}, J. and {Tricco}, T.~S. and {Nixon}, C. and {Toupin}, S. and {Pettitt}, A. and {Chan}, C. and {Mentiplay}, D.and {Laibe}, Guillaume and {Glover}, Simon and {Dobbs}, C. and {Nealon}, R. and {Liptai}, D. and {Worpel}, H. and {Bonnerot}, C. and {Dipierro}, G. and {Ballabio}, G. and {Ragusa}, E. and {Federrath}, C. and {Iaconi}, Roberto and {Reichardt}, Thomas and {Forgan}, Duncan and {Hutchison}, M. and {Constantino}, T. and {Ayliffe}, B. and {Hirsh}, K. and {Lodato}, G.},
	doi = {10.1017/pasa.2018.25},
	eid = {e031},
	journal = {Publications of the Astronomical Society of Australia},
	month = Sep,
	pages = {e031},
	primaryclass = {astro-ph.IM},
	title = {{Phantom: A Smoothed Particle Hydrodynamics and Magnetohydrodynamics Code for Astrophysics}},
	volume = {35},
	year = 2018}

@article{choi2016,
	adsurl = {https://ui.adsabs.harvard.edu/abs/2016ApJ...823..102C},
	archiveprefix = {arXiv},
	author = {{Choi}, Jieun and {Dotter}, Aaron and {Conroy}, Charlie and {Cantiello}, Matteo and {Paxton}, Bill and {Johnson}, Benjamin D.},
	doi = {10.3847/0004-637X/823/2/102},
	eid = {102},
	eprint = {1604.08592},
	journal = {\apj},
	month = jun,
	number = {2},
	pages = {102},
	primaryclass = {astro-ph.SR},
	title = {{Mesa Isochrones and Stellar Tracks (MIST). I. Solar-scaled Models}},
	volume = {823},
	year = 2016}

@article{haworth2020,
	adsurl = {https://ui.adsabs.harvard.edu/abs/2020MNRAS.494.4130H},
	archiveprefix = {arXiv},
	author = {{Haworth}, Thomas J. and {Cadman}, James and {Meru}, Farzana and {Hall}, Cassandra and {Albertini}, Emma and {Forgan}, Duncan and {Rice}, Ken and {Owen}, James E.},
	doi = {10.1093/mnras/staa883},
	eprint = {2001.06225},
	journal = {\mnras},
	month = may,
	number = {3},
	pages = {4130-4148},
	primaryclass = {astro-ph.SR},
	title = {{Massive discs around low-mass stars}},
	volume = {494},
	year = 2020}

@article{kratter2011,
	adsurl = {https://ui.adsabs.harvard.edu/abs/2011ApJ...740....1K},
	archiveprefix = {arXiv},
	author = {{Kratter}, Kaitlin M. and {Murray-Clay}, Ruth A.},
	doi = {10.1088/0004-637X/740/1/1},
	eid = {1},
	eprint = {1107.0728},
	journal = {\apj},
	month = oct,
	number = {1},
	pages = {1},
	primaryclass = {astro-ph.SR},
	title = {{Fragment Production and Survival in Irradiated Disks: A Comprehensive Cooling Criterion}},
	volume = {740},
	year = 2011}

@article{kratter2010,
	adsurl = {https://ui.adsabs.harvard.edu/abs/2010ApJ...710.1375K},
	archiveprefix = {arXiv},
	author = {{Kratter}, Kaitlin M. and {Murray-Clay}, Ruth A. and {Youdin}, Andrew N.},
	doi = {10.1088/0004-637X/710/2/1375},
	eprint = {0909.2644},
	journal = {\apj},
	month = feb,
	number = {2},
	pages = {1375-1386},
	primaryclass = {astro-ph.EP},
	title = {{The Runts of the Litter: Why Planets Formed Through Gravitational Instability Can Only Be Failed Binary Stars}},
	volume = {710},
	year = 2010}

@article{meru2012,
	adsurl = {https://ui.adsabs.harvard.edu/abs/2012MNRAS.427.2022M},
	archiveprefix = {arXiv},
	author = {{Meru}, Farzana and {Bate}, Matthew R.},
	doi = {10.1111/j.1365-2966.2012.22035.x},
	eprint = {1209.1107},
	journal = {\mnras},
	month = dec,
	number = {3},
	pages = {2022-2046},
	primaryclass = {astro-ph.EP},
	title = {{On the convergence of the critical cooling time-scale for the fragmentation of self-gravitating discs}},
	volume = {427},
	year = 2012}

@article{lodato2011,
	adsurl = {https://ui.adsabs.harvard.edu/abs/2011MNRAS.413.2735L},
	archiveprefix = {arXiv},
	author = {{Lodato}, Giuseppe and {Clarke}, C.~J.},
	doi = {10.1111/j.1365-2966.2011.18344.x},
	eprint = {1101.2448},
	journal = {\mnras},
	month = jun,
	number = {4},
	pages = {2735-2740},
	primaryclass = {astro-ph.SR},
	title = {{Resolution requirements for smoothed particle hydrodynamics simulations of self-gravitating accretion discs}},
	volume = {413},
	year = 2011}

@article{nelson2006,
	adsurl = {https://ui.adsabs.harvard.edu/abs/2006MNRAS.373.1039N},
	archiveprefix = {arXiv},
	author = {{Nelson}, Andrew F.},
	doi = {10.1111/j.1365-2966.2006.11119.x},
	eprint = {astro-ph/0609493},
	journal = {\mnras},
	month = dec,
	number = {3},
	pages = {1039-1073},
	primaryclass = {astro-ph},
	title = {{Numerical requirements for simulations of self-gravitating and non-self-gravitating discs}},
	volume = {373},
	year = 2006}

@article{rafikov2009,
	adsurl = {https://ui.adsabs.harvard.edu/abs/2009ApJ...704..281R},
	archiveprefix = {arXiv},
	author = {{Rafikov}, Roman R.},
	doi = {10.1088/0004-637X/704/1/281},
	eprint = {0901.4739},
	journal = {\apj},
	month = oct,
	number = {1},
	pages = {281-291},
	primaryclass = {astro-ph.EP},
	title = {{Properties of Gravitoturbulent Accretion Disks}},
	volume = {704},
	year = 2009}

@article{clarke2009,
	adsurl = {https://ui.adsabs.harvard.edu/abs/2009MNRAS.396.1066C},
	archiveprefix = {arXiv},
	author = {{Clarke}, C.~J.},
	doi = {10.1111/j.1365-2966.2009.14774.x},
	eprint = {0904.3549},
	journal = {\mnras},
	month = jun,
	number = {2},
	pages = {1066-1074},
	primaryclass = {astro-ph.SR},
	title = {{Pseudo-viscous modelling of self-gravitating discs and the formation of low mass ratio binaries}},
	volume = {396},
	year = 2009}

@article{young2024,
	adsurl = {https://ui.adsabs.harvard.edu/abs/2024MNRAS.tmp.1254Y},
	archiveprefix = {arXiv},
	author = {{Young}, Alison K. and {Celeste}, Maggie and {Booth}, Richard A. and {Rice}, Ken and {Koval}, Adam and {Carter}, Ethan and {Stamatellos}, Dimitris},
	doi = {10.1093/mnras/stae1249},
	eprint = {2405.05762},
	journal = {\mnras},
	month = may,
	primaryclass = {astro-ph.SR},
	title = {{Introducing two improved methods for approximating radiative cooling in hydrodynamical simulations of accretion discs}},
	year = 2024}

@article{kratter2016,
	adsurl = {https://ui.adsabs.harvard.edu/abs/2016ARA&A..54..271K},
	archiveprefix = {arXiv},
	author = {{Kratter}, Kaitlin and {Lodato}, Giuseppe},
	doi = {10.1146/annurev-astro-081915-023307},
	eprint = {1603.01280},
	journal = {\araa},
	month = sep,
	pages = {271-311},
	primaryclass = {astro-ph.SR},
	title = {{Gravitational Instabilities in Circumstellar Disks}},
	volume = {54},
	year = 2016}

@article{paczynski1978,
	adsurl = {https://ui.adsabs.harvard.edu/abs/1978AcA....28...91P},
	author = {{Paczynski}, B.},
	journal = {\actaa},
	month = jan,
	pages = {91-109},
	title = {{A model of selfgravitating accretion disk.}},
	volume = {28},
	year = 1978}

@article{toomre1964,
	adsurl = {https://ui.adsabs.harvard.edu/abs/1964ApJ...139.1217T},
	author = {{Toomre}, A.},
	doi = {10.1086/147861},
	journal = {\apj},
	month = may,
	pages = {1217-1238},
	title = {{On the gravitational stability of a disk of stars.}},
	volume = {139},
	year = 1964}

@article{mercer2018,
	adsurl = {https://ui.adsabs.harvard.edu/abs/2018MNRAS.478.3478M},
	archiveprefix = {arXiv},
	author = {{Mercer}, A. and {Stamatellos}, D. and {Dunhill}, A.},
	doi = {10.1093/mnras/sty1290},
	eprint = {1805.09568},
	journal = {\mnras},
	month = aug,
	number = {3},
	pages = {3478-3493},
	primaryclass = {astro-ph.EP},
	title = {{Efficient radiative transfer techniques in hydrodynamic simulations}},
	volume = {478},
	year = 2018}

@article{lombardi2015,
	adsurl = {https://ui.adsabs.harvard.edu/abs/2015MNRAS.447...25L},
	archiveprefix = {arXiv},
	author = {{Lombardi}, James C. and {McInally}, William G. and {Faber}, Joshua A.},
	doi = {10.1093/mnras/stu2432},
	eprint = {1411.3678},
	journal = {\mnras},
	month = feb,
	number = {1},
	pages = {25-35},
	primaryclass = {astro-ph.IM},
	title = {{An efficient radiative cooling approximation for use in hydrodynamic simulations}},
	volume = {447},
	year = 2015}

@article{stamatellos2007,
	adsurl = {https://ui.adsabs.harvard.edu/abs/2007A&A...475...37S},
	archiveprefix = {arXiv},
	author = {{Stamatellos}, D. and {Whitworth}, A.~P. and {Bisbas}, T. and {Goodwin}, S.},
	doi = {10.1051/0004-6361:20077373},
	eprint = {0705.0127},
	journal = {\aap},
	month = nov,
	number = {1},
	pages = {37-49},
	primaryclass = {astro-ph},
	title = {{Radiative transfer and the energy equation in SPH simulations of star formation}},
	volume = {475},
	year = 2007}

@article{gammie2001,
	adsurl = {https://ui.adsabs.harvard.edu/abs/2001ApJ...553..174G},
	archiveprefix = {arXiv},
	author = {{Gammie}, Charles F.},
	doi = {10.1086/320631},
	eprint = {astro-ph/0101501},
	journal = {\apj},
	month = may,
	number = {1},
	pages = {174-183},
	primaryclass = {astro-ph},
	title = {{Nonlinear Outcome of Gravitational Instability in Cooling, Gaseous Disks}},
	volume = {553},
	year = 2001}

@article{longarini2023,
	adsurl = {https://ui.adsabs.harvard.edu/abs/2023MNRAS.519.2017L},
	archiveprefix = {arXiv},
	author = {{Longarini}, Cristiano and {Lodato}, Giuseppe and {Bertin}, Giuseppe and {Armitage}, Philip J.},
	doi = {10.1093/mnras/stac3653},
	eid = {arXiv:2212.04986},
	eprint = {2212.04986},
	journal = {\mnras},
	month = feb,
	number = {2},
	pages = {2017-2029},
	primaryclass = {astro-ph.EP},
	title = {{The role of the drag force in the gravitational stability of dusty planet forming disc - I. Analytical theory}},
	volume = {519},
	year = 2023}

@article{rice2006,
	adsurl = {https://ui.adsabs.harvard.edu/abs/2006MNRAS.372L...9R},
	archiveprefix = {arXiv},
	author = {{Rice}, W.~K.~M. and {Lodato}, G. and {Pringle}, J.~E. and {Armitage}, P.~J. and {Bonnell}, I.~A.},
	doi = {10.1111/j.1745-3933.2006.00215.x},
	eprint = {astro-ph/0607268},
	journal = {\mnras},
	month = oct,
	number = {1},
	pages = {L9-L13},
	primaryclass = {astro-ph},
	title = {{Planetesimal formation via fragmentation in self-gravitating protoplanetary discs}},
	volume = {372},
	year = 2006}

@article{segura-cox2020,
	adsurl = {https://ui.adsabs.harvard.edu/abs/2020Natur.586..228S},
	archiveprefix = {arXiv},
	author = {{Segura-Cox}, Dominique M. and {Schmiedeke}, Anika and {Pineda}, Jaime E. and {Stephens}, Ian W. and {Fern{\'a}ndez-L{\'o}pez}, Manuel and {Looney}, Leslie W. and {Caselli}, Paola and {Li}, Zhi-Yun and {Mundy}, Lee G. and {Kwon}, Woojin and {Harris}, Robert J.},
	doi = {10.1038/s41586-020-2779-6},
	eprint = {2010.03657},
	journal = {\nat},
	month = oct,
	number = {7828},
	pages = {228-231},
	primaryclass = {astro-ph.EP},
	title = {{Four annular structures in a protostellar disk less than 500,000 years old}},
	volume = {586},
	year = 2020}

@article{nixon2018,
	adsurl = {https://ui.adsabs.harvard.edu/abs/2018MNRAS.477.3273N},
	archiveprefix = {arXiv},
	author = {{Nixon}, C.~J. and {King}, A.~R. and {Pringle}, J.~E.},
	doi = {10.1093/mnras/sty593},
	eprint = {1803.04417},
	journal = {\mnras},
	month = jul,
	number = {3},
	pages = {3273-3278},
	primaryclass = {astro-ph.EP},
	title = {{The Maximum Mass Solar Nebula and the early formation of planets}},
	volume = {477},
	year = 2018}

@article{cadman2020a,
	adsurl = {https://ui.adsabs.harvard.edu/abs/2020MNRAS.492.5041C},
	archiveprefix = {arXiv},
	author = {{Cadman}, James and {Rice}, Ken and {Hall}, Cassandra and {Haworth}, Thomas J. and {Biller}, Beth},
	doi = {10.1093/mnras/staa187},
	eprint = {2001.06224},
	journal = {\mnras},
	month = mar,
	number = {4},
	pages = {5041-5051},
	primaryclass = {astro-ph.EP},
	title = {{Fragmentation favoured in discs around higher mass stars}},
	volume = {492},
	year = 2020}



\appendix

\section{Setting initial conditions}
\label{sec:app_init}

We need to be careful that the choice of initial conditions does not have undue influence on the outcomes of the simulations. To verify that the results presented are robust, we repeat a sample of the simulations after taking steps to ensure the initial discs are stable and settled. The simulations chosen are given in table \ref{tab:extrasims}.

\begin{table}
    \centering
    \begin{tabular}{c|c|c|c|c}
     ID & Outcome &  $M_*$ (\solmass{}) & $R_{\rm out}$ (au) &  $M_{\rm d}/M_*$\\
     \hline
      A & axisymmetric & 1.0 & 200 & 0.4 \\ 
      S & spiral & 0.5 & 50 & 0.3 \\
      F & fragments & 0.7 & 100 & 0.5
    \end{tabular}
    \caption{Details of the simulations that were repeated with slight changes to initial conditions and initially evolved for 2 ORPs with $\beta$-cooling.}
    \label{tab:extrasims}
\end{table}

The aim of altering the initial conditions was to obtain an initial disc that was as stable as possible, i.e. to minimise initial spreading inwards and outwards due to artificially abrupt pressure gradients and to reduce any initial transient over-densities. The surface density power law exponent was reduced to $p=0.75$ and density was tapered more gently in the outer disc from $r_c = 0.8 R_{\rm out}$.
The disc was initialised with $5.2\times 10^5$ particles to allow for increased initial accretion. After 2 ORPs, there were $\approx 4$~per cent more particles in the discs than at the same point in the original simulations.
The discs were then allowed to evolve for 2 ORPs with fixed $\beta_{\rm cool} = 20$, so that any initial perturbations were not amplified by changes in optical depth. From there, the simulations were evolved as before with the modified Lombardi cooling.

\begin{figure}
    \centering
    \includegraphics[width=0.8\linewidth]{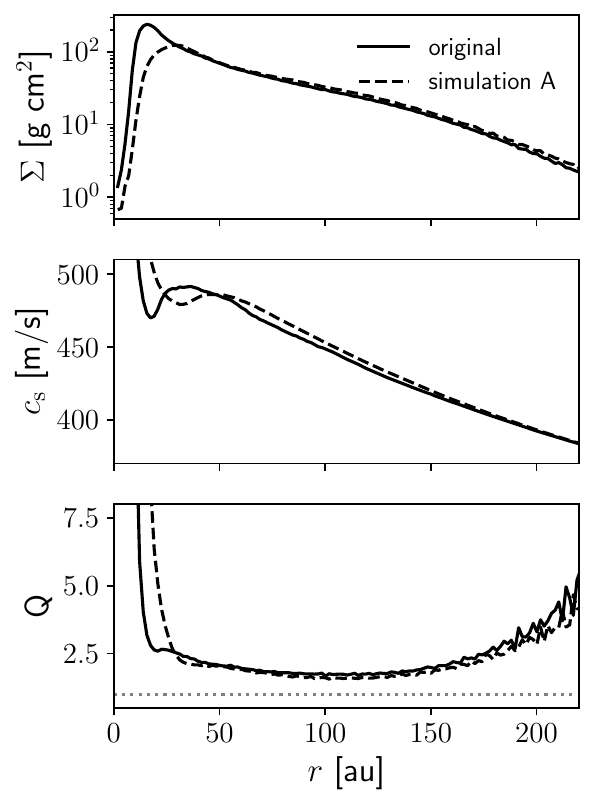}
    \caption{Azimuthally averaged surface density, sound speed and $Q$ profiles for snapshots taken after 4 ORPs, for simulation A (axisymmetric) and the corresponding original simulation from the main paper.}
    \label{fig:simA}
\end{figure}

\begin{figure}
    \centering
    \includegraphics[width=0.9\linewidth,trim= 0cm 6cm 0cm 0cm,clip]{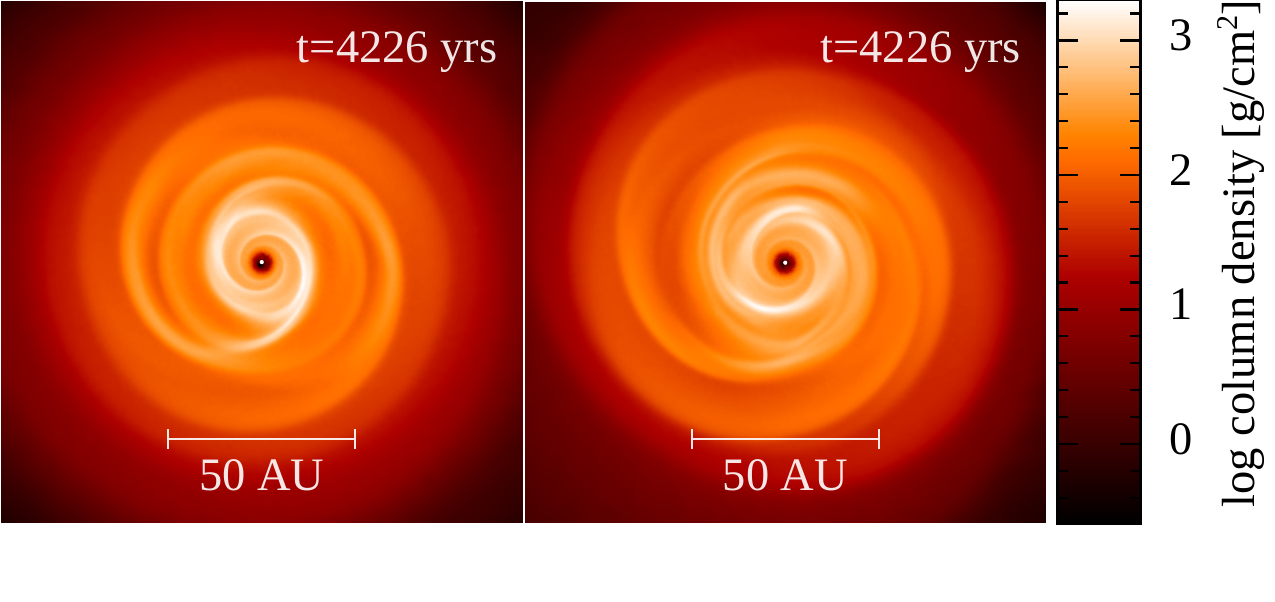}
    \\
    \includegraphics[width=0.9\linewidth,trim= 0cm 6cm 0cm 0cm,clip]{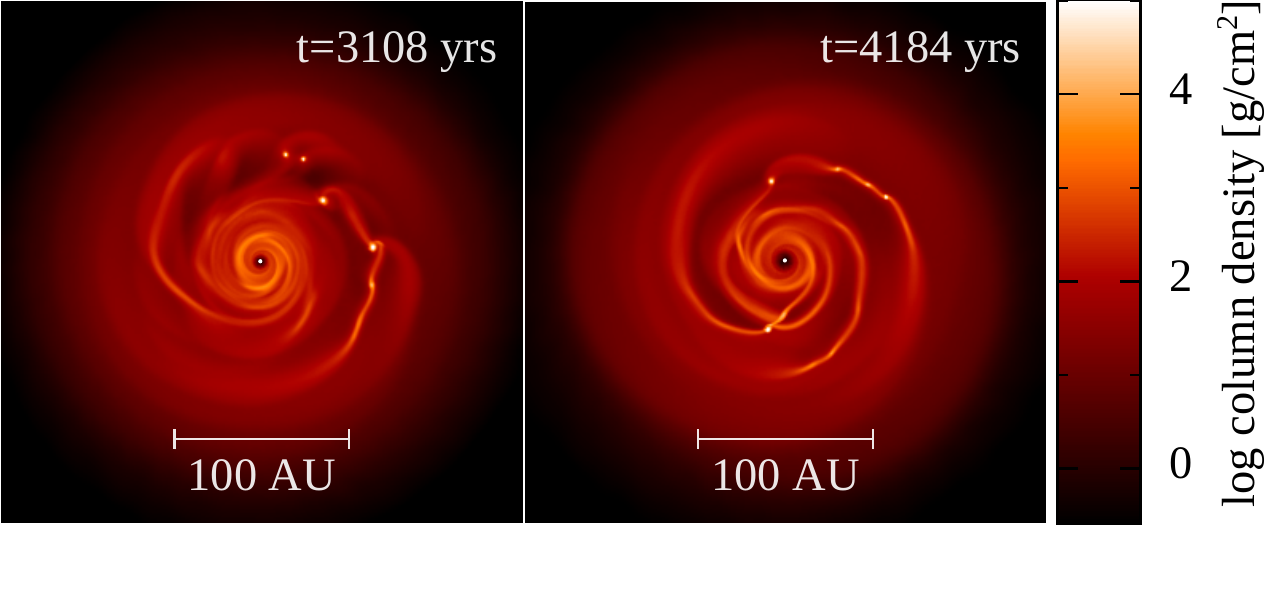}
    \caption{Top right: new simulation S; Bottom right: new simulation F. The counterpart snapshots from the original simulations are in the left hand panels. The structures are broadly very similar in the new simulations.}
    \label{fig:newICsims}
\end{figure}

The outcome of the new simulations is the same as the original simulations. Fig.~\ref{fig:simA} shows azimuthally averaged properties of the disc after 4 ORPs in the original simulation compared to simulation A. The density peak is much shallower at the inner edge and beyond 50~au the sound speed and Toomre Q are similar. Simulations S and F are shown in Fig.~\ref{fig:newICsims} with the original simulations in the left hand panels. We can see that the structures are largely similar, with simulation S displaying a stable spiral structure as before, and simulation F producing collapsing fragments. We can therefore conclude that the results presented in this paper are not driven by the initial conditions.


\section{Resolution}
\label{sec:resolution}
In this work we chose to model the discs with $5\times10^5$ particles so that a parameter study was feasible with the available compute resources. This is sufficient to resolve the Toomre mass \citep{nelson2006}, which reaches a minimum of $\sim$~\solmass{0.002}, with $>27$ SPH particles. As an additional check we repeat some simulations at a higher resolution of $10^6$ SPH particles. Fig.~\ref{fig:resolution_frag} shows that the disc evolution is very similar for a high mass disc. Fragmentation occurs slightly sooner with the higher resolution, but the development of spiral waves and clumps proceeds in a similar manner.

In Fig.~\ref{fig:resolution_lines}, we compare the surface density, $Q$, vertical resolution, and mid-plane optical depth for two discs modelled with increased resolution. Increasing the resolution has a small effect, causing a reduced optical depth since the mid plane is better resolved. The vertical resolution is just sufficient with $\langle h\rangle/H \approx 0.25$ for the regions of interest.
We also tested the effects of changing the lower limit of $\alpha_{\rm SPH}$ used by the viscosity switch for the $R=50$~au disc (left panels). The long term evolution of all four simulations was similar, only the time at which spiral structures developed differed. The optical depths and cooling rates are broadly similar with differences due mainly to the speed of evolution.

\begin{figure}
    \centering
    \includegraphics[trim= 0cm 7cm 0cm 0cm,clip,width=1\linewidth]{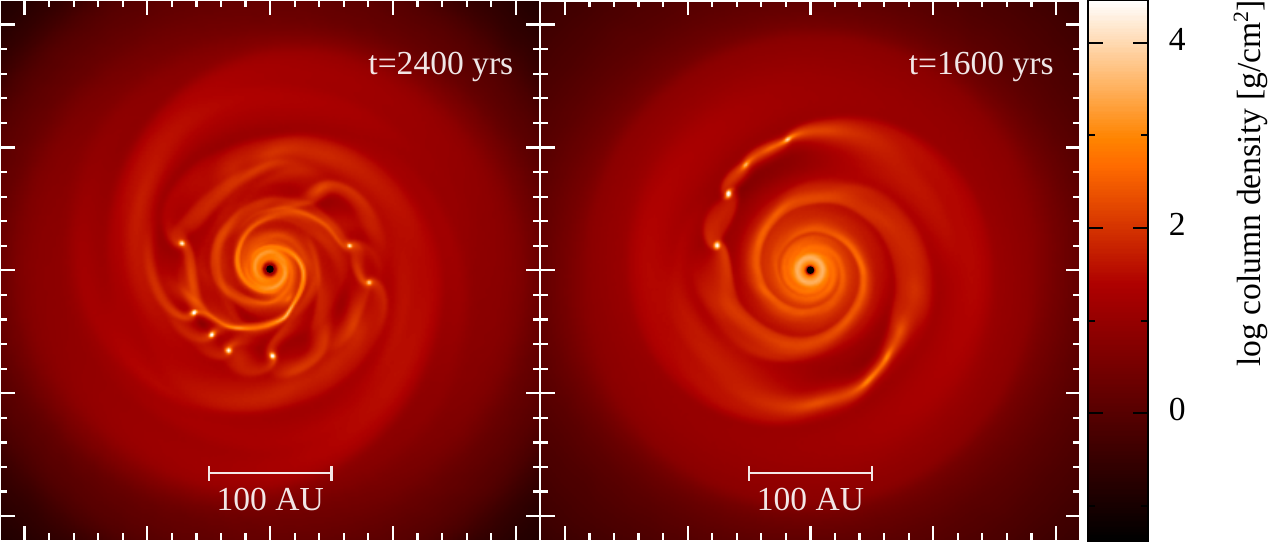}
    \includegraphics[trim= 0cm 7cm 0cm 0cm,clip,width=1\linewidth]{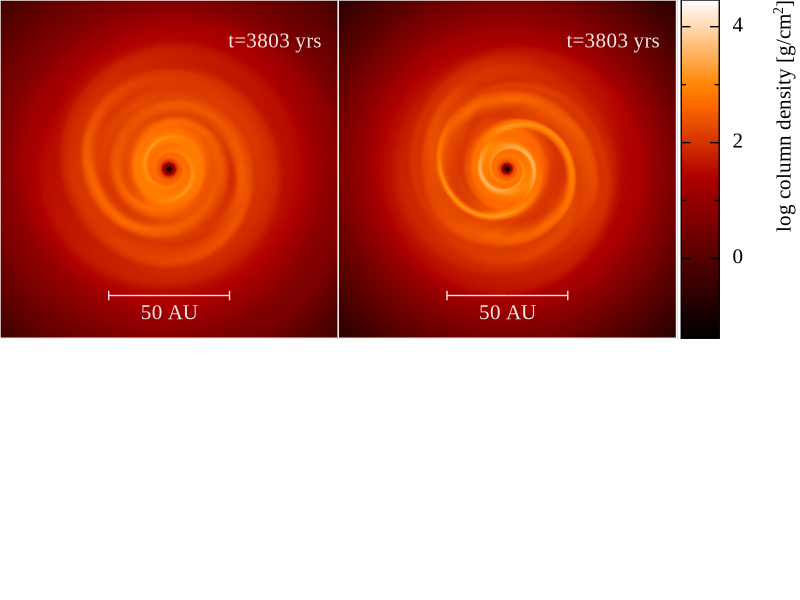}
    \caption{Comparison of the onset of fragmentation in a simulation using $5\times 10^5$ particles (left) and $10^6$ particles (right). Upper panels: model had $M_*=$~\solmass{0.5}, $R_{\rm out} = 200$~au and $M_{\rm d}/M_*=1$. Lower panels: $M_*=$~\solmass{0.7}, $R_{\rm out} = 50$~au and $M_{\rm d}/M_*=0.3$.  Fragmentation occurs slightly sooner with the higher resolution, still within a fraction of an ORP. In the lower panels, the structure is broadly similar, however it took around twice as long for the $m=2$ spiral to form in the high resolution simulation.}
    \label{fig:resolution_frag}
\end{figure}

\begin{figure}
    \centering
    \includegraphics[width=0.8\linewidth]{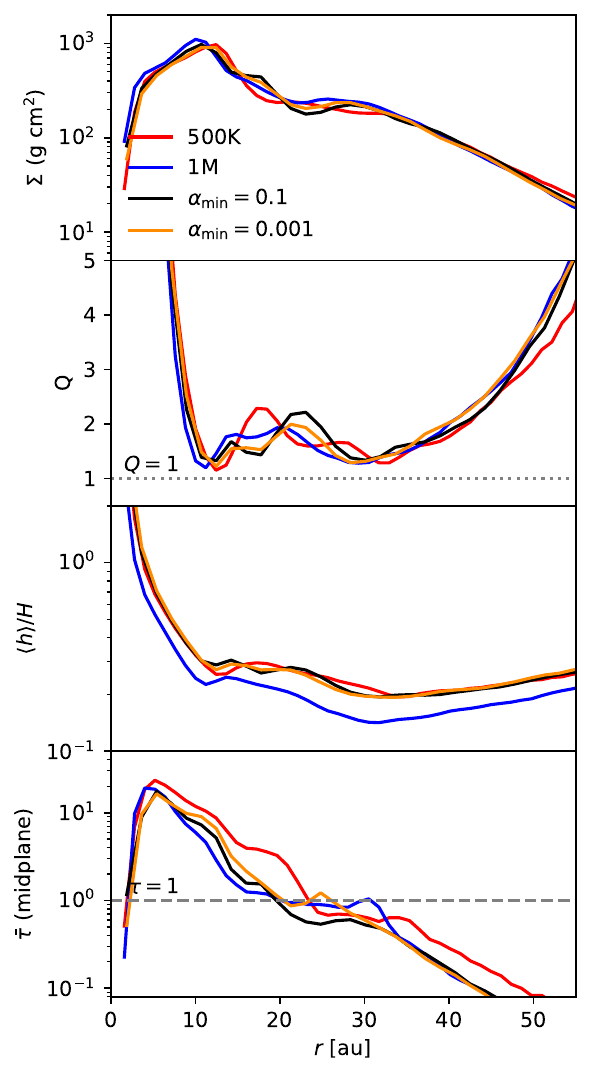}
    \caption{Comparison of simulations with standard resolution ($5\times 10^5$ particles) and high resolution. $M_*=$~\solmass{0.7}, $R_{\rm out} = 50$~au and $M_*/M_{\rm d}=0.3$ (See also bottom panels of Fig.~\ref{fig:resolution_frag}). These snapshots were taken at 10 ORPs. Also shown are the outcomes of simulations with $5\times 10^5$ particles and a raised and reduced minimum value of $\alpha_{\rm SPH}$. }
    \label{fig:resolution_lines}
\end{figure}

\section{Comparison with Haworth et al. (2020)}
\label{sec:app_haworth}
While the method for calculating the radiative cooling rate is identical, the temperature of the disc due to stellar heating is different. In our simulations, the minimum disc temperature is determined by the stellar irradiation accounting for some attenuation of the radiation incident on the disc surface using equation \ref{eq:discmintemp}. \citet{haworth2020} did not apply any attenuation so the disc temperatures are effectively estimated in the optically thin limit, resulting in higher temperatures despite employing identical values of stellar luminosity (Table \ref{tab:luminosity}). We therefore now compare the evolution of discs under the Stamatellos method in this work with that of \citet{haworth2020} to determine to what extent the difference in minimum temperature estimates affects the outcome. The results are shown in Fig.~\ref{fig:Stam_outcomes}, with our simulations in black and those of \citet{haworth2020} in red. We note that the latter work does not distinguish between faint and large-scale spirals.

We find evolution of the discs to be similar. In the top panel of Fig.~\ref{fig:Stam_outcomes} it is apparent that the gravitational instability becomes active at slightly lower values of $M_{\rm d}/M_*$ for 50~au discs, driving the formation of spiral arms. This can be explained by the lower disc temperatures in the new simulations due to the attenuation of the stellar irradiation. For the 200~au discs (Fig.~\ref{fig:Stam_outcomes}, lower panel) we find that fragmentation occurs at slightly higher disc masses. Since the mid-plane temperature is lower in our simulations, the gravitational instability is able to self-regulate effectively to a higher $M_{\rm d}/M_*$ by locally increasing the temperature enough to prevent fragmentation.

Both approaches to calculating the minimum disc temperatures are approximations that represent rough upper and lower limits. Comparing with Fig.~\ref{fig:gridoutcomes}, it's clear that the method of estimating the radiative cooling rate has a more significant effect on the disc evolution than this difference in minimum disc temperature.

\begin{figure}
    \centering
    \includegraphics[width=0.8\linewidth]{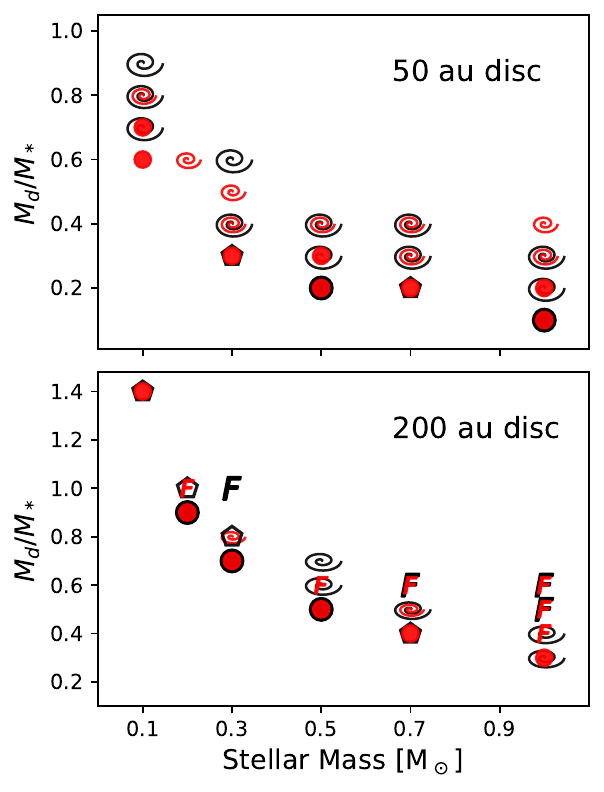}
    \caption{A comparison of the evolution of discs using the MIST stellar luminosity from \citet{haworth2020} (red icons) with our code using the Stamatellos method and same luminosities (black icons). 'F' indicates fragmentation, spirals indicate large-scale spiral structures, pentagons indicate faint spirals (only for our simulations), and circles indicate axisymmetric discs.}
    \label{fig:Stam_outcomes}
\end{figure}


\bsp	
\label{lastpage}
\end{document}